\documentclass[times,twocolumn,final]{elsarticle}

\usepackage{medima}
\usepackage{framed,multirow}
\usepackage{amssymb}
\usepackage{latexsym}
\usepackage{xcolor}
\usepackage{lineno,hyperref}
\modulolinenumbers[5]
\usepackage{amsmath}
\usepackage{verbatim}
\usepackage{multirow}
\usepackage{color,graphicx}
\usepackage{nicematrix}
\usepackage{arydshln}
\usepackage{hyperref}
\usepackage{url}
\usepackage{footnote}
\usepackage{bm}
\usepackage{cite}
\usepackage{subfigure}
\usepackage{verbatim}
\usepackage{ulem}
\usepackage{graphicx}
\usepackage{pgfplots}
\usepackage{cancel}

\providecommand{\Leirefeq}[1]{Equation (\ref{#1})}
\providecommand{\Leireftb}[1]{Table~\ref{#1}}
\providecommand{\Leireffig}[1]{Fig.~\ref{#1}}
\providecommand{\Leireffigure}[1]{Figure~\ref{#1}}

\usepackage{enumitem}

\journal{Medical Image Analysis}

\begin{document}

\verso{Lei Li \textit{et~al.}}

\begin{frontmatter}

\title{Personalized Topology-Informed Localization of Standard 12-Lead ECG Electrode Placement from Incomplete Cardiac MRIs for Efficient Cardiac Digital Twins}  

\author[label1,label2,label3]{Lei Li} 
\ead[url]{lei.sky.li@soton.ac.uk}
\author[label4]{Hannah Smith} 
\author[label1]{Yilin Lyu} 
\author[label4]{Julia Camps} 
\author[label5]{Shuang Qian} 
\author[label4]{Blanca Rodriguez} 
\author[label2]{Abhirup Banerjee} 
\author[label2]{Vicente Grau} 
 
\address[label1]{School of Electronics \& Computer Science, University of Southampton, Southampton, UK}
\address[label2]{Department of Engineering Science, University of Oxford, Oxford, UK}
\address[label3]{Department of Biomedical Engineering, National University of Singapore, Singapore}
\address[label4]{Department of Computer Science, University of Oxford, Oxford, UK}
\address[label5]{School of Biomedical Engineering and Imaging Sciences, Kings College London, London, UK}

\received{19 Aug 2024}

\begin{abstract}
Cardiac digital twins (CDTs) offer personalized \textit{in-silico} cardiac representations for the inference of multi-scale properties tied to cardiac mechanisms. 
The creation of CDTs requires precise information about the electrode position on the torso, especially for the personalized electrocardiogram (ECG) calibration.
However, current studies commonly rely on additional acquisition of torso imaging and manual/semi-automatic methods for ECG electrode localization.
In this study, we propose a novel and efficient topology-informed model to fully automatically extract personalized ECG standard electrode locations from 2D clinically standard cardiac MRIs. 
Specifically, we obtain the sparse torso contours from the cardiac MRIs and then localize the standard electrodes of 12-lead ECG from the contours.
Cardiac MRIs aim at imaging of the heart instead of the torso, leading to incomplete torso geometry within the imaging.
To tackle the missing topology, we incorporate the electrodes as a subset of the keypoints, which can be explicitly aligned with the 3D torso topology.
The experimental results demonstrate that the proposed model outperforms the time-consuming conventional model projection-based method in terms of accuracy (Euclidean distance: $1.24 \pm 0.293$ cm vs. $1.48 \pm 0.362$ cm) and efficiency ($2$~s vs. $30$-$35$~min).
We further demonstrate the effectiveness of using the detected electrodes for \textit{in-silico} ECG simulation, highlighting their potential for creating accurate and efficient CDT models.
The code is available at \url{https://github.com/lileitech/12lead_ECG_electrode_localizer}.
\end{abstract}


\end{frontmatter}


\section{Introduction}


\begin{figure}[t]\center
 \includegraphics[width=0.5\textwidth]{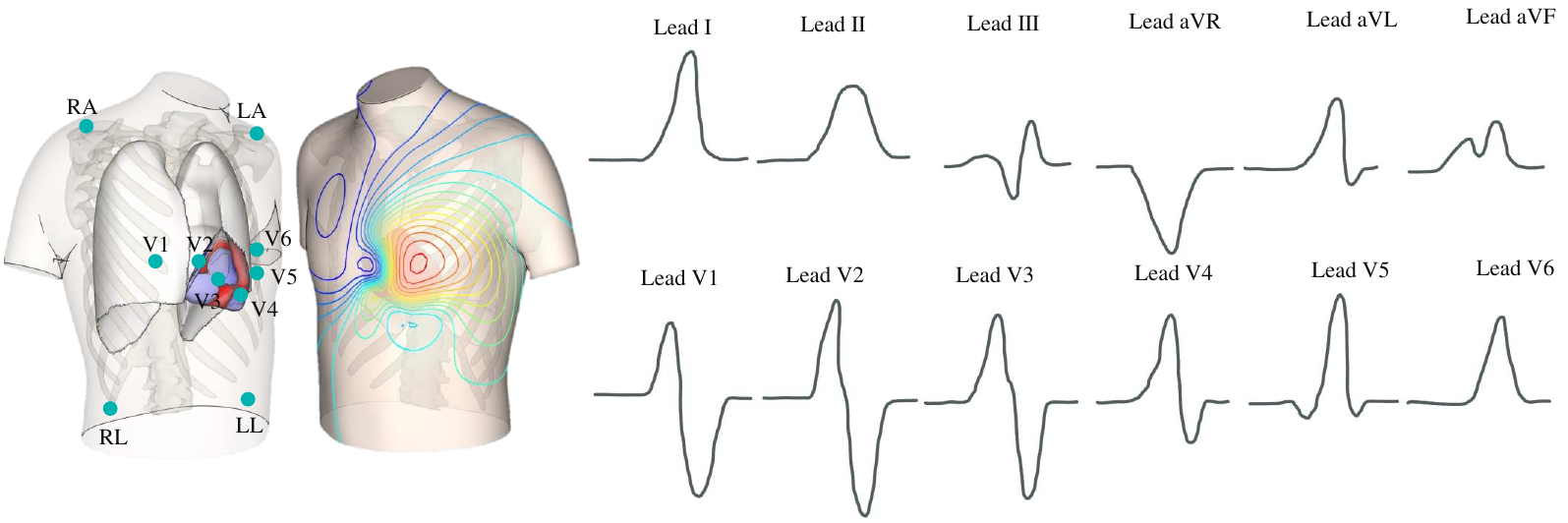}\\[-2ex]
   \caption{Illustration of the setup of the torso with electrodes (labeled with cyan-blue dots) and simulated 12-lead ECG.} 
\label{fig:intro:digital_twins}
\end{figure}

Cardiac digital twins (CDTs), also known as virtual heart models, aim to replicate the anatomical and physiological characteristics of an individual heart \citep{journal/MedIA/gillette2021,journal/MedIA/camps2024}. 
Their ability to simulate and predict personalized responses to interventions makes them invaluable tools for advancing precision medicine and improving patient outcomes in cardiovascular care \citep{journal/Nature_Comm/arevalo2016,journal/Nature_BME/boyle2019,journal/Nature_BME/prakosa2018}.
The development of CDTs becomes possible due to the availability of patient-specific information from electrical data (e.g. electrocardiogram, ECG) and imaging data \citep{journal/TMI/li2024}.
ECG, a widely utilized tool for recording the electrical activity within the heart, is commonly performed with 12 leads.
The standard 12-lead ECG utilizes 10 electrodes to measure body surface potential, inferring cardiac potential based on heart and torso anatomy and their relative positions \citep{journal/JE/roy2020,journal/FiP/gillette2022}. 
The variations in electrode positions can influence the duration and amplitudes of recorded ECG waveforms \citep{journal/MBEC/kania2014,journal/JE/rjoob2020}. 
Therefore, the electrode positions serve as a key factor in accurately capturing the relationship between the electrical activity of the heart and the surrounding torso anatomy \citep{journal/JE/van2014}.
Accurate electrode positioning typically requires specialized devices, such as custom ECG systems with electrode localizers or specialized photography, which are generally unavailable in routine clinical practice \citep{journal/MedIA/gillette2021,journal/TMI/ghanem2003,journal/JE/roudijk2021}. 
Furthermore, given the challenges of localizing standard limb electrodes via medical imaging, the Mason-Likar (M-L) modification of the standard 12-lead ECG has been developed. 
It relocates limb electrodes (RA, LA, RL, LL) from the arms and legs to the torso while retaining the precordial leads (V1–V6) in their standard chest positions \citep{journal/JE/man2008}, as shown in \Leireffig{fig:intro:digital_twins}. 
Although this altered placement may alter ECG morphology, a study by \citet{journal/JE/man2008} demonstrated that standard 12-lead ECGs can be effectively reconstructed from M-L configurations using specific transformation matrices.
Consequently, current studies on creating digital twins primarily employ standard torso-based M-L electrode locations identified through medical imaging \citep{journal/MedIA/camps2021,book/ITSCE/loewe2022,journal/FiP/gillette2022,journal/medRxiv/qian2023,journal/MedIA/camps2024,journal/MedIA/camps2024a,journal/JRSI/salvador2024,journal/TMI/li2024}.



\begin{figure*}[t]\center
 \includegraphics[width=0.8\textwidth]{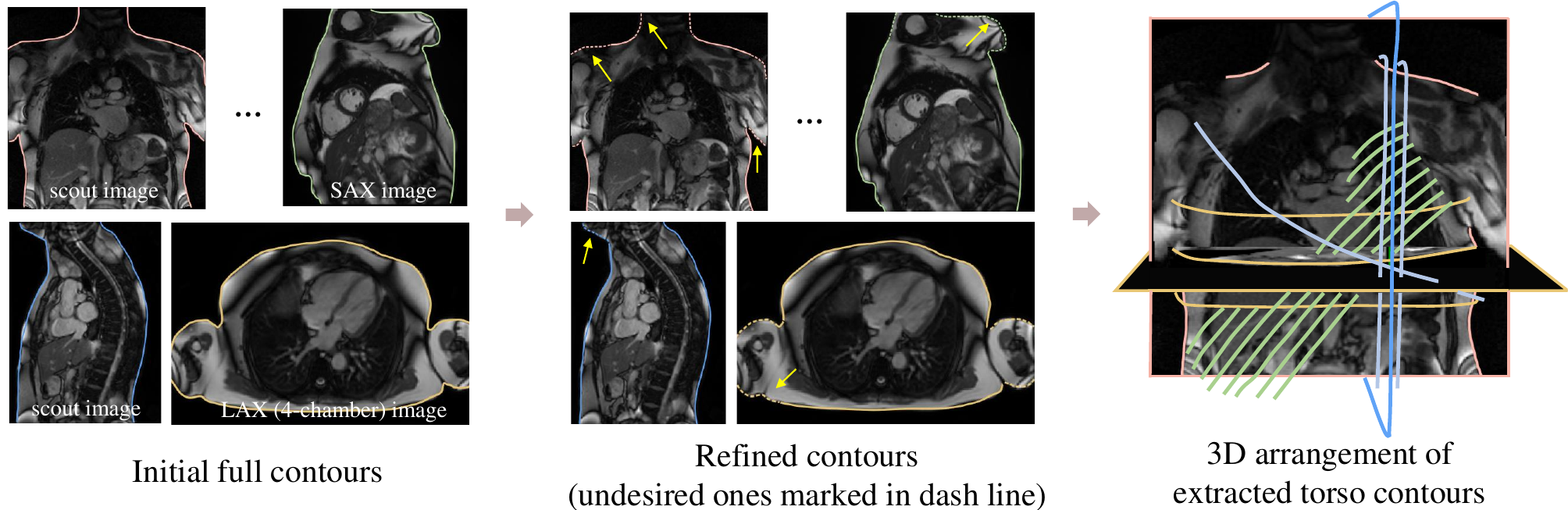}\\[-2ex]
   \caption{Visualization of extracted full and refined contours from cardiac MRIs and the 3D arrangement of the refined torso contours (only partially visualized here). Here, the acquired cardiac MRIs include scout images covering a larger range of the torso, typically used as localizers, and heart-focused slices in various views, including long-axis (LAX) views (2-, 3-, and 4-chamber) and a stack of short-axis (SAX) views.}
\label{fig:method:torso_contour}
\end{figure*}

Cardiac imaging, typically obtained through magnetic resonance imaging (MRI) or computed tomography (CT) scans, offers detailed insights into the anatomical, morphological, and functional features of the heart \citep{journal/MedIA/li2023}. 
These imaging techniques have proved invaluable for reconstructing cardiac anatomy and pinpointing pathology locations. 
However, standard cardiac imaging poses challenges in reconstructing torso geometry and localizing electrodes \citep{journal/arXiv/smith2023,conf/EMBC/smith2022}. 
Since these imaging views primarily focus on the heart, there is limited information available on the contours of the sections of the torso distant from the heart, leading to an incomplete representation of the torso, as presented in \Leireffig{fig:method:torso_contour}.
Furthermore, unwanted features such as the head, neck, and arms may be present in the images, which can hinder and negatively impact torso reconstruction \citep{conf/CiC/gillette2015}.
Also, common MRI artifacts often manifest as shadow regions, particularly above the shoulder or near the waist.

Therefore, additional torso images are usually required for heart-torso geometry modeling and electrode localization \citep{journal/Nature_Medicine/ramanathan2004,journal/MedIA/gillette2021,journal/FiP/gillette2022}.
However, obtaining full torso geometry necessitates additional scanning time, leading to heightened patient discomfort due to prolonged breath-holding and an escalation in the cost of MRI data acquisition.
Few studies attempted to reconstruct torso geometry directly from cardiac MRI and then localize the electrodes, which was challenging due to partial torso shape information \citep{conf/MICCAI/zacur2017,conf/EMBC/smith2022}.
These studies commonly integrated statistical shape models (SSMs) into the reconstruction process to impose prior anatomical constraints about the torso geometry, which however were often time-consuming.
This is because SSM-based electrode detection typically involves iterative optimization of 3D torso geometry by aligning the individual shape with a mean torso model.

In this work, we develop a novel pipeline to efficiently localize the standard torso electrodes from 2D clinically acquired cardiac MRIs.
Note that the cardiac imaging setup includes scout images that partially cover the torso to serve as localizers for the subsequent cardiac cine imaging.
The proposed method incorporates topology information about the torso into the pipeline to guide the electrode localization directly from incomplete contours, which are extracted from cardiac MRIs. 
This is achieved by introducing the keypoint detection and surface skeleton-assisted point cloud completion (PCN) modules.
As shown in \Leireffig{fig:intro:digital_twins}, ECG electrodes are positioned at anatomically significant locations to capture cardiac electrical activity, particularly the limb electrodes, which serve as key landmarks for the torso geometry (specifically the front side of the torso). 
To employ the spatial relationship between the electrodes and torso geometry, we adopt the electrodes as a subset of the keypoints used for torso reconstruction. 
Therefore, an end-to-end learning framework is created for ECG electrode localization and 3D torso reconstruction from incomplete contours.
To the best of our knowledge, this is the first fully automatic deep learning based torso electrode localization work, addressing the challenges of incomplete information through the incorporation of topology information.
The main contributions of this work include:
\begin{enumerate}[label=\roman*.]
  \item We develop a novel topology-informed method for estimating standard torso electrode location from 2D clinically acquired cardiac MRIs.
  \item We convert the challenging electrode localization problem into the keypoint detection task, which can be explicitly informed by the 3D torso topology.
  \item We reconstruct 3D torso geometry from keypoints via a surface skeleton-assisted PCN, to further employ the spatial relationship between torso and keypoints (electrodes).  
  \item We prove the feasibility of accurately inferring personalized electrode location from cardiac MRIs to create an efficient CDT platform for precision medicine.
\end{enumerate}

\section{Related Work}

\subsection{Cardiac Digital Twins: Bridging Precision Cardiology}

CDT technology aims to create a personalized virtual cardiac representations based on patient-specific data, such as imaging, ECG, etc \citep{journal/TMI/li2024,journal/MedIA/camps2024a}. 
It has shown great promise in the personalized treatments for cardiac disease, such as the stratification of arrhythmia risk \citep{journal/Nature_Comm/arevalo2016} and the ablation guidance of persistent atrial fibrillation \citep{journal/Nature_BME/boyle2019}.
The creation of a digital heart usually involves two stages, namely anatomical twinning and functional twinning \citep{journal/MedIA/gillette2021,journal/TMI/li2024}.
For the anatomical twining, one needs to perform the cardiac segmentation, 3D heart-torso geometry reconstruction, and the modeling of relevant pathology. 
This process is complicated by the variability in cardiac anatomy among different individuals and by imaging artifacts and noise.
The functional twining mainly includes the ECG simulation and calibration via solving the ECG inverse problem, which is inherently ill-posed \citep{journal/RBME/li2024}.  
This process is further complicated by the limitations of ECG recordings, which are sparse, noisy, and subject to substantial uncertainties.
Furthermore, the torso geometry and electrode localization accuracy can influence the inverse solution, as the electrophysiological (EP) measurement is generally performed on the body surface \citep{conf/CinC/zemzemi2015,journal/medRxiv/qian2023}.

\subsection{Torso Reconstruction and Electrode Localization}

Most existing electrode localization studies focus on identifying standard M-L electrode positions on the torso for ECG simulation \citep{journal/FiP/minchole2019,journal/MedIA/gillette2021,journal/JRSI/salvador2024}.
These methods typically rely on semi-automatic techniques and often require additional torso imaging for precise 3D torso reconstruction \citep{journal/FiP/minchole2019,journal/MedIA/gillette2021}.
For example, \citet{journal/JRSI/salvador2024} created heart-torso models from full torso CT scan using semi-automatic methods, i.e., threshold- and region-growing-based segmentation methods, and then electrodes were identified on the torso.
\citet{journal/TBME/giffard2016} reconstructed the torso mesh from CT images and identified the electrodes for 12-lead ECG from body surface potential (BSP) mapping electrodes based on the standard torso ECG placement.
Instead of acquiring torso imaging, several studies reconstructed torso geometry from cardiac MRIs and then localized the standard torso electrodes on the 3D torso \citep{conf/MICCAI/zacur2017,conf/EMBC/smith2022}.
To consider the torso heterogeneity, \citet{conf/MICCAI/zacur2017} expanded the torso representation to encompass ribs and lungs and then virtually placed electrodes on it to synthesize a 12-lead ECG.
Nevertheless, including various organs in computational studies remains controversial, as some research indicates minimal contributions of organ impedances on BSP \citep{journal/TBME/ghanem2003}, while others emphasize significant effects attributable to uncertainties in impedance values \citep{journal/TBME/keller2010}.

Instead of localizing the virtual electrodes on the torso, a few studies directly employed specialized CT/ MRI or photography to localize the actual electrodes \citep{journal/TMI/ghanem2003,journal/JE/perez2018,journal/JE/roudijk2021}.
For example, \citet{journal/Nature_Medicine/ramanathan2004} manually segmented slice by slice to obtain heart-torso geometry from CT images, which are combined with X-ray-opaque Ag/AgCl electrodes, allowing for simultaneous acquisition of body-surface electrode locations.
\citet{journal/MedIA/gillette2021} employed MRI-compatible electrodes that were placed before image acquisition to obtain the electrode location.
Then, they segmented the torso from the full torso MRI using threshold-based filters with boolean operations in Seg3D\footnote{https://www.sci.utah.edu/cibc-software/seg3d.html} and then generated anatomical torso meshes using an image-based unstructured mesh generation technique.
\citet{journal/TMI/ghanem2003} employed 2D photography (SONY DSC-S85) and point reconstruction techniques to determine ECG electrode positions, calibrating with a known object and matching 3D coordinates of identified electrodes in sequential photographs.
Nowadays, 3D cameras such as Intel Real Sense D435 can be employed to directly capture 3D photos, which can be fitted to a standard thoracic model and then the electrode positions can be segmented based on these 3D photos \citep{journal/JE/roudijk2021}.
While these approaches utilize specialized imaging for precise electrode localization of the actual ECG electrode placement, our work focuses on detecting the standard electrode positions of 12-lead ECG using routine available imaging, addressing a distinct challenge in clinical practice.

\begin{figure*}[t]\center
 \includegraphics[width=1\textwidth]{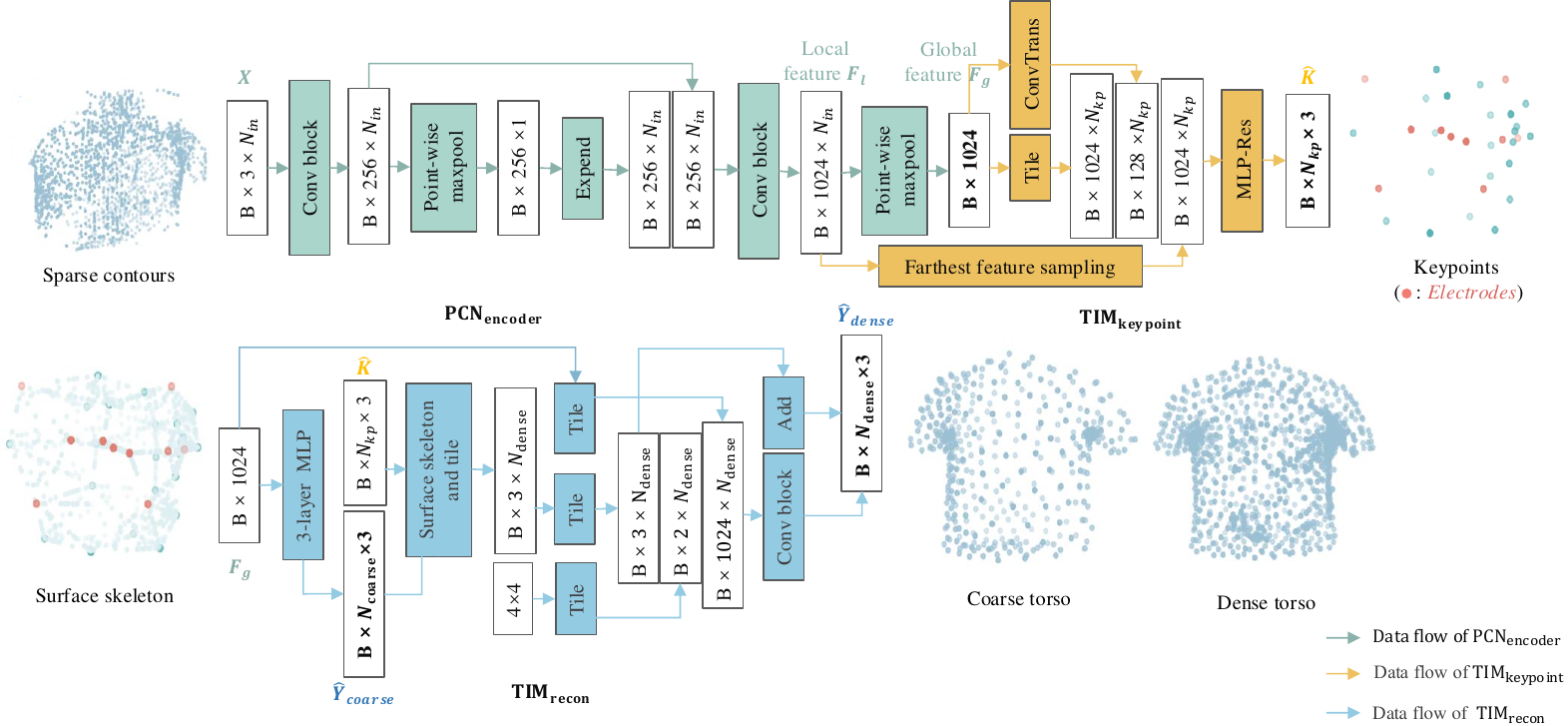}\\[-2ex]
   \caption{Topology-informed model (TIM) for 12-lead ECG electrode detection and torso geometry reconstruction from sparse and incomplete torso contours. Here, the electrodes are labeled via red dots within the predicted keypoints, which are then converted into surface skeleton to guide the 3D torso reconstruction. Note that the diagram takes 32 keypoints as an example. PCN: point completion model. }
\label{fig:method:topology-informed model}
\end{figure*}

\subsection{Incomplete Data Processing}

In real-world applications, the data may be partially missing, corrupted by noise, or exhibit other forms of incompleteness \citep{journal/PAA/aste2015}.
In the context of medical imaging, incomplete data refers to scenarios where the images lack coverage of specific regions of interest (ROIs), or when the dataset of the target patient includes a series of imaging sequences or views, with some of these sequences or views being absent.
It may arise due to various factors such as imaging limitations, out-of-range measurements, or challenges in capturing specific ROIs. 
For example, many head CT images in clinical practice are obtained without conducting a complete head scan, aiming to minimize radiation exposure \citep{journal/MedIA/wan2023}.
The subjects with myocardial infarction may not have full sequences for the pathology analysis, such as missing late gadolinium enhancement (LGE) MRI or mapping MRI scans, due to patient-specific conditions \citep{journal/MedIA/qiu2023}.
The long-axis MRI data may be missing for 3D cardiac geometry reconstruction, resulting in incomplete shape information especially at the apex and base of the heart \citep{journal/CBM/sander2023}.
Therefore, it is essential to develop models with the capability of managing incomplete imaging information in a flexible manner.
Traditional methods of processing incomplete imaging data often involve incorporating prior knowledge, such as SSM, to help reconstruct or complete missing information \citep{conf/MICCAI/zacur2017,conf/EMBC/smith2022}.
These methods require expensive optimizations during inference and thus are impractical for real-time applications.
To mitigate this issue, \citet{journal/MedIA/wan2023} embedded the SSM into the model training to guide the learning process for solutions that align with the prior knowledge.
Furthermore, data-driven image synthesis has been widely used for processing incomplete imaging data \citep{conf/JP/wang2020,journal/Circulation/zhang2021}. 
Integrating information from non-imaging data can be a valuable strategy for completing missing knowledge within imaging data \citep{journal/TMI/li2024}.
Recently, several works used a deep learning based parameterized model to directly map partial input to a complete shape \citep{conf/3DV/yuan2018,conf/CVPR/tang2022,journal/MedIA/beetz2023}.
However, these approaches do not explicitly learn the topology of the complete shape, resulting in less robust predictions, particularly in missing regions.

\section{Methodology}

Figure~\ref{fig:method:topology-informed model} presents the pipeline of the proposed topology informed model. 
The whole pipeline includes five parts: torso contour extraction, point feature extraction, keypoint (electrode) localization, surface skeleton generation, and 3D torso reconstruction.
In detail, the input partial point clouds (resampled from torso contours) are firstly fed into an encoder to learn both local and global features. 
Then, we localize torso keypoints, where we identify a subset of the keypoints as electrodes.
At the same time, the global features are employed to generate coarse torso representation, which is subsequently used to guide the keypoints to create corresponding surface skeletons.
At the end, the surface skeletons are fed into the refinement part to generate the dense torso. 

\subsection{Torso Contour Extraction from Multi-View Cardiac MRIs} \label{method:contour}

The cardiac MRIs are acquired using standard clinical protocols not designed for torso detection, often including structures and artifacts that interfere with reconstruction, and sometimes missing parts of the region of interest.
The reader is referred to Sec. \ref{exp:data} for detailed information on the MRI data used in the experiments.
Therefore, we use a two-step automated framework that comprises a segmentation network followed by a two-channel refinement network for the torso contour extraction \citep{conf/EMBC/smith2022}.
Specifically, we firstly employ a U-Net based torso segmentation model to extract all visible sections of the body and then convert it into full contours. 
Next, a U-Net based refinement model is utilized to eliminate contour sections influenced by undesired anatomical features (e.g., head, neck, and connected arm sections) and any boundaries caused by MRI artifacts.
Note that the input of the refinement model include both the image and the corresponding full contour obtained in the first step.
Both models are optimized by maximizing the Dice score between the predicted mask and ground truth mask. 
As shown in \Leireffig{fig:method:torso_contour}, the extracted torso contours can be arranged together into a sparse torso representation based on their 3D world coordinates.
Nonetheless, the torso representation is sparse and incomplete, and thus some of the electrodes may not be described in the sparse contours.

\begin{figure*}[t]\center
 \includegraphics[width=0.9\textwidth]{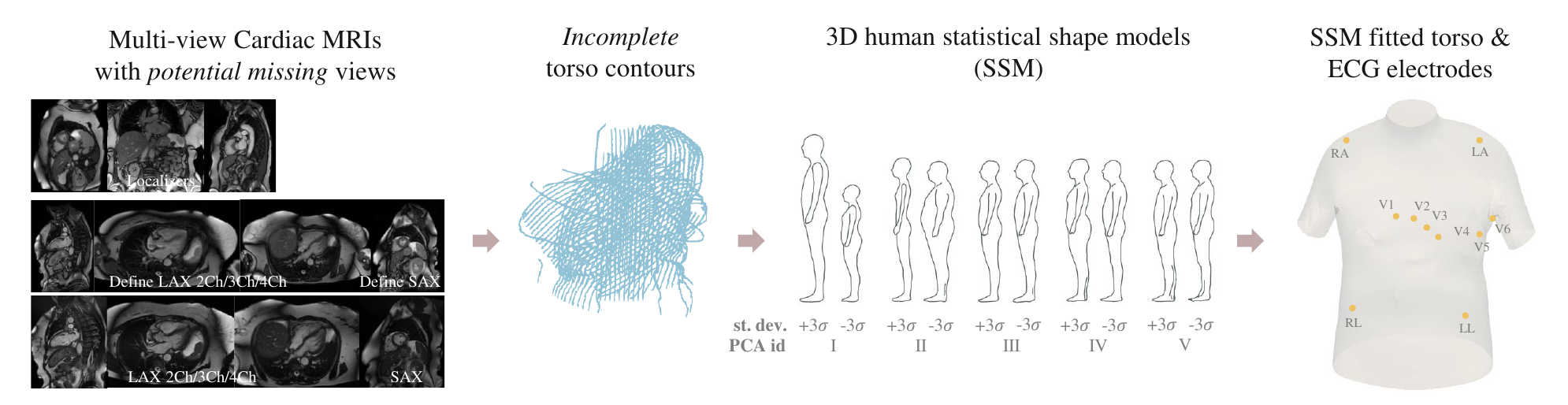}\\[-2ex]
   \caption{ Statistical shape model (SSM) based 3D torso mesh reconstruction and electrode localization from multi-view 2D cardiac MRIs.}
\label{fig:data:SSM}
\end{figure*}

\subsection{Torso Keypoint Detector for Electrode Localization} \label{method:keypoint}

To localize the electrodes from the sparse and incomplete torso contours, we consider the electrodes as a subset of the keypoints.
Keypoints in 3D geometry are important points that highlight key features or structures, which are essential for accurately reconstructing the 3D shape.  
As we know, standard torso ECG electrodes are located at specific locations on the front side of the torso.
Therefore, the electrode localization task can be converted into a keypoint detection task, wherein we leverage the inherent relationship between electrodes and torso geometry.
Given a set of incomplete point clouds $X = \{x_i \,|\, i = 1, \ldots , N_\text{in}\} \in \mathbb{R}^{N_{in} \times 3}$, our objective is to predict $N_{kp}$ keypoints denoted as $K = \{ K_j \,|\, j = 1, \ldots , N_\text{electrode}, \ldots, N_{kp} \} \in \mathbb{R}^{N_{kp} \times 3}$, which may not be present in $X$. 
Among these keypoints, $K_\text{electrode} = \{ K_j \,|\, j = 1, \ldots , N_\text{electrode} \} \in \mathbb{R}^{N_\text{electrode} \times 3}$ specifically represents electrodes for 12-lead ECG ($N_\text{electrode}=10$). 
As \Leireffig{fig:method:topology-informed model} shows, we employ an extended version of PointNet as a backbone encoder to extract both local and global features \citep{conf/3DV/yuan2018}, denoted as $F_{l}$ and $F_{g}$. 
Then, we employ the farthest feature sampling to extract keypoints, which are then converted into surface skeleton via surface interpolation \citep{conf/CVPR/tang2022}.

The keypoints consist of all electrodes along with additional complementary points needed to fully represent the topology of the entire torso.
Therefore, they can be partially supervised by minimizing the mean absolute error (MAE) between the ground truth and predicted electrodes (denoted as $\hat{K}_\text{electrode}$),
\begin{equation}
	\mathcal{L}_\text{electrode} = \mathcal{L}^\text{MAE}(\hat{K}_\text{electrode}, K_\text{electrode}).
\end{equation}
To embed the torso topology as a constraint, we further introduce a loss for keypoint detection via calculating the Chamfer Distance (CD) between the predicted keypoints $\hat{K}$ and the torso topology, 
\begin{equation}
	\mathcal{L}_\text{keypoint} = \mathcal{L}^\text{CD}_\text{keypoint}(\hat{K}, Y_\text{topology}),
\end{equation} 
where $Y_\text{topology}$ can be generated via farthest point sampling (FPS), which can iteratively select the farthest points from the already selected points to obtained uniformly distributed keypoints across the 3D torso.
Here, \(\mathcal{L}_\text{electrode}\) serves as a strong constraint for keypoint detection, while \(\mathcal{L}_\text{keypoint}\) imposes a soft constraint on the overall spatial arrangement of the localized electrodes.


\subsection{Keypoint Skeleton-Assisted 3D Torso Reconstruction} \label{method:geometry}

To reconstruct the 3D torso geometry, the detected keypoints are converted into a surface skeleton, as an intermediate representation between keypoints and final torso geometry.
The reconstruction process initiates by utilizing global features as input for a three-layer multilayer perceptron (MLP), facilitating the reconstruction of the coarse torso geometry. 
Subsequently, we employ a surface-skeleton generation algorithm \citep{conf/CVPR/tang2022} to establish the torso skeleton from keypoints along the coarse torso surface.
The skeleton comprises a mixture of curves and triangular surfaces that dynamically adapt to the underlying 3D geometry.
The dense torso geometry is then reconstructed combining both topological and geometric information of torso from the surface skeleton and the coarse geometry.
The geometry reconstruction loss is defined as
\begin{equation}
	\mathcal{L}_\text{rec} = \mathcal{L}_\text{torso}^{CD}(\hat{Y}_\text{coarse}, Y_\text{coarse}) + \beta \mathcal{L}_\text{torso}^{CD}(\hat{Y}_\text{dense}, Y_\text{dense}),
\end{equation}
where $\beta$ is the weight term between the geometry loss for coarse and dense torso point clouds, and $\mathcal{L}_\text{torso}^{CD}$ is defined as the CD between the predicted $\hat{Y}$ and ground truth torso geometries $Y$.
Therefore, the total loss of the TIM is defined by combining all the related losses mentioned above,
\begin{equation}
	\mathcal{L} = \mathcal{L}_\text{electrode} + \lambda_\text{keypoint} \mathcal{L}_\text{keypoint} + \lambda_\text{rec} \mathcal{L}_\text{rec},
\end{equation}
where $\lambda_\text{keypoint}$ and $\lambda_\text{rec}$ are balancing parameters.

\begin{figure}[t]\center
 \includegraphics[width=0.5\textwidth]{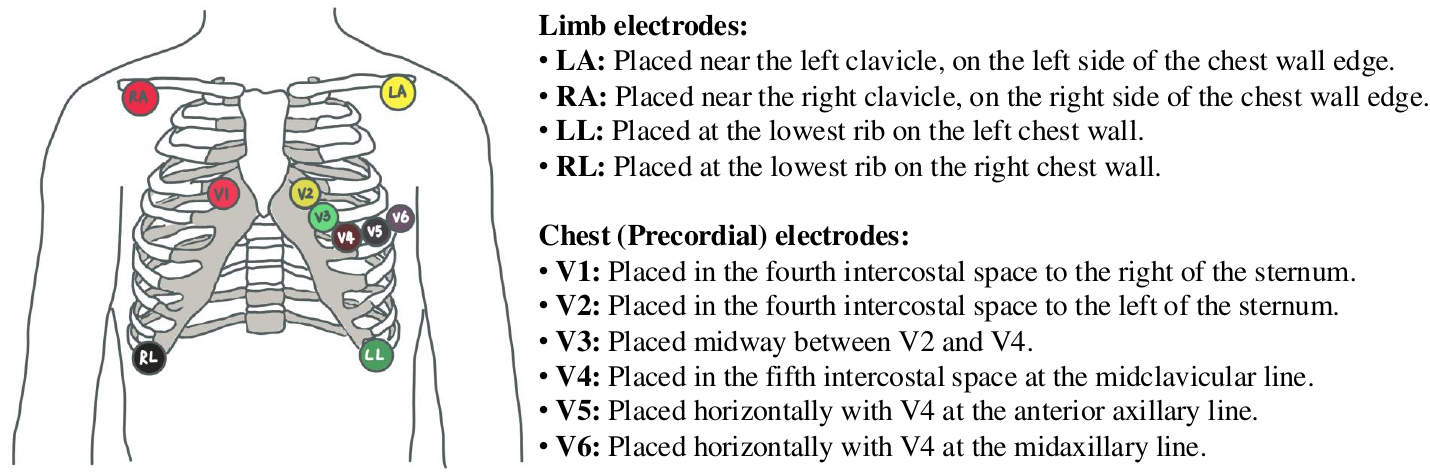}\\[-2ex]
   \caption{The standard placement of torso electrodes for 12-lead ECG.}
\label{fig:data:electrode_definition}
\end{figure}

\section{Experiments and Results}

\subsection{Data Acquisition and Pre-Processing} \label{exp:data}

We employed multi-view cardiac MRIs from 200 subjects as part of the UK Biobank study (application number 40161).
The cardiac MRIs were acquired on 1.5 Tesla scanner (MAGNETOM Aera, Siemens Healthcare, Erlangen, Germany) \citep{journal/JCMR/petersen2016}. 
This scanner is equipped with 48 receiver channels, a gradient system of 45 mT/m and 200 T/m/s, an 18-channel anterior body surface coil, and a 12-element integrated spine coil with ECG gating for cardiac synchronization. 
The acquisition process begins with scout images, also known as localizers, providing partial coverage of the torso in sagittal, transverse, and coronal planes \citep{journal/JCMR/kramer2020}. 
These images guide the placement and alignment of focused cardiac scans, which include three long-axis (LAX) cine sequences (2-, 3-, and 4-chamber views) and a stack of short-axis (SAX) balanced steady-state free precession (bSSFP) cines.
Here, we employed both heart-focused bSSFP cine images at the end-diastolic (ED) phase and the localizers to extract the torso contours.
The subjects, aged 40-70 years, had a 1:1 gender ratio and an average body mass index (BMI) of $26.6 \pm 4.04$ $kg/m^2$.
The dataset was randomly divided into 120 training subjects, 20 validation subjects, and 60 test subjects.

\subsection{Gold Standard and Evaluation}

For evaluation, we compared the predicted electrodes with the gold standard.
The exact electrode positions of 12-lead ECG were not available in the UK Biobank dataset, so we employed a conventional SSM-based method \citep{journal/arXiv/smith2023} to generate the personalized 3D torso meshes and then manually selected electrodes on the reconstructed torso.
As \Leireffig{fig:data:SSM} shows, the SSM generated from a cluster of human body shapes \citep{journal/PR/pishchulin2017} was applied on the sparse torso contours within a 3D space.
The SSM was first translated into the subject-specific coordinate system by aligning the cardiac centers based on the intersections of cine SAX and LAX planes with the sparse contours. 
Subsequently, the initial torso shape was deformed to fit the contours by minimizing the distances between them, producing the final 3D torso mesh.
The generated torso meshes obtained the average contour-to-surface distance of $0.546 \pm 0.208$ cm.
The M-L electrodes (LA, RA, LL, RL, and V1-V6) were manually placed on the torso referring to their standard positions (see \Leireffig{fig:data:electrode_definition}) by a well-trained PhD student using ParaView\footnote{https://www.paraview.org} and checked by a senior expert. 
This torso reconstruction and manual electrode localization for each subject typically required about 30-35 min and 3 min, respectively.
Although the electrodes are synthetic, the process closely simulates clinical practice, where clinicians manually place electrodes on the body surface of patient according to standardized guidelines. 
Minor placement variations, common in clinical environments, are reflected in our synthetic data, although we did not intentionally introduce such variations in our training or test sets.

For torso geometry reconstruction, we employed CD to assess the alignment between the predicted torso geometry and the reconstructed one by SSM. 
The electrode localization errors were reported in terms of mean of Euclidean distance (ED) between the detected electrode coordinates and the corresponding ground truth.
Furthermore, we performed an \textit{in-silico} evaluation by comparing simulated ECG morphology based on predicted electrodes and ground truth electrodes.
Note that this work primarily focuses on electrode localization, which serves as a crucial input for EP simulation in CDTs. 
Torso reconstruction, while important, is not a necessary step in our pipeline during inference. 
Therefore, we mainly report the localization results in the following experiments to highlight the critical role of accurate electrode positioning in ensuring the fidelity of EP simulations.

\subsection{Implementation} \label{exp:implementation}

The framework was implemented in PyTorch, running on a computer with 2.1~GHz 13th Gen Intel(R) Core(TM) i7-13700 CPU and an NVIDIA GeForce RTX 3070. 
We used the Adam optimizer to update the network parameters with weight decay of 1e-3. 
The batch size ($B$) was set to 6, and the initial learning rate was set to 1e-4 and multiplied by 0.5 every 9,000 iterations. 
The balancing parameters were set empirically as follows: 
$\beta=5$,
$\lambda_\text{keypoint}=0.05$, 
and $\lambda_\text{rec}=0.05$.
The input torso contour has been converted into a point cloud and uniformly resampled to 2048 points. 
To mitigate random resampling effects, data augmentation was performed by repeating the random resampling $N_{rr}$ times, and the effect of $N_{rr}$ was studied in Sec. \ref{exp:parameter study:rr}.
The reconstructed coarse and dense torso geometries, represented as point clouds, consisted of $N_\text{coarse}$ (=1024) and $N_\text{dense}$ (=4096) points, respectively.
The number of keypoints $N_{kp}$ was set to 32 initially and was investigated in Sec. \ref{exp:parameter study:kp}, while the number of points in torso topology $Y_\text{topology}$ was set to 128.
The training of the model took about 1 hour (120 epochs in total), while the inference of the networks required about 2~s to process each subject.

\subsection{Parameter Studies}

\subsubsection{Effect of the Number of Random Resampling} \label{exp:parameter study:rr}

\begin{figure}[t]\center
 \includegraphics[width=0.42\textwidth]{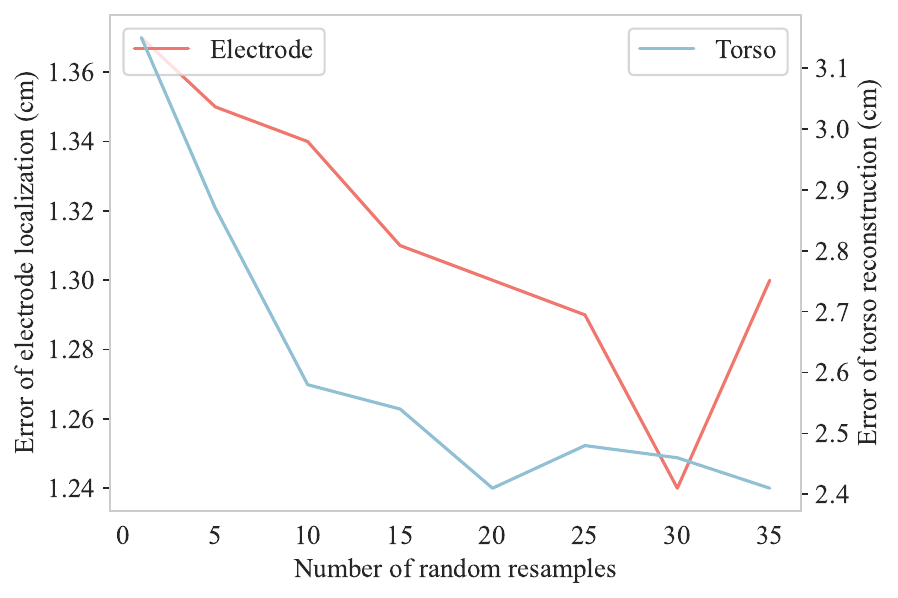}\\[-2ex] 
   \caption{Performance of the proposed method against different number of random resampling. } 
\label{fig:exp:rr}
\end{figure}

To investigate the effect of random resampling, we compared the performance of the proposed method with different numbers of random resampling $N_{rr}$.
Here, we set $N_{rr}$ ranging from 1 to 45 (number of keypoints $N_{kp}$ = 32), as presented in \Leireffig{fig:exp:rr}.
The electrode error $ED_\text{electrode}$ steadily decreases as $N_{rr}$ increases from 1 to 30, with the lowest error achieved at $N_{rr}$ = 30.
This indicates that increasing the number of resamples helps in better approximating the electrode positions, likely due to the increased variability and robustness in the sampling process that averages out random errors.
For the torso reconstruction error, the performance becomes stable once $N_{rr}$ reaches 20. 
This suggests that beyond this point, additional resampling does not significantly improve the reconstruction accuracy. 
This plateau effect is expected as the resampling process sufficiently captures the variability in the torso contours, allowing the method to consistently reconstruct the torso geometry accurately.
Overall, the results indicate that an optimal balance is achieved with $N_{rr} = 30$ for the proposed method.

\begin{table} [t] \center
    \caption{
     Summary of the quantitative evaluation results of the proposed method with different numbers of keypoints. CD: Chamfer distance; ED: Euclidean distance. 
     }
\label{tb:exp:kp}
{\small
\begin{tabular}{  l| l l *{3}{@{\ \,} l }}
\hline
Num of keypoints        & $CD_\text{torso}$ (cm)  & $ED_\text{electrode}$ (cm) \\
\hline
10 (only electrodes)    & $ 2.46 \pm 0.287 $      & $ 1.35 \pm 0.349 $ \\
16                      & $ 2.43 \pm 0.203 $      & $ 1.28 \pm 0.292 $ \\ 
32                      & $ 2.46 \pm 0.306 $      & $ 1.24 \pm 0.291 $ \\ 
64                      & $ 2.29 \pm 0.235 $      & $ 1.24 \pm 0.293 $ \\ 
128                     & $ 2.31 \pm 0.308 $      & $ 1.28 \pm 0.314 $ \\
\hline
\end{tabular} }\\
\end{table}

\subsubsection{Effect of the Number of Keypoints} \label{exp:parameter study:kp}

In this study, we first compared the results of the proposed scheme using different numbers of keypoints $N_{kp}$ to investigate its effect on the proposed method.
Here, we set the value ranging from the baseline one, i.e., only 10 electrodes, to 128, as presented in \Leireftb{tb:exp:kp}. 
One can see that the best performance was obtained when the number of keypoints was set to 64.
This indicates that only electrodes are not enough to present the complete torso topology as a constraint for both accurate electrode localization and torso reconstruction.
The ratio between the number of electrodes and that of keypoints should be balanced to achieve optimal performance.
In the following experiments, $N_{kp}$ was set to 64 for the proposed method.

\subsection{Comparison and Ablation Study} \label{exp:compare}

\begin{table} [t] \center
    \caption{
     Summary of the quantitative evaluation results of different methods. 
     }
\label{tb:exp:results}
{\small
\begin{tabular}{  l| l l *{3}{@{\ \,} l }}
\hline
Method                   & $CD_\text{torso}$ (cm)  & $ED_\text{electrode}$ (cm) \\
\hline
SSM projection               & N/A                     & $ 1.48 \pm 0.362 $ \\
PCN$_\text{encoder}$ + 3FC   & N/A                     & $ 2.00 \pm 0.661 $ \\ 
TIM \textit{w/o} kp          & $ 2.38 \pm 0.306 $      & $ 1.54 \pm 0.446 $ \\
TIM \textit{w/o} recon       & N/A                     & $ 1.32 \pm 0.338 $ \\
TIM                          & $ 2.29 \pm 0.235 $      & $ 1.24 \pm 0.293 $ \\
\hline
\end{tabular} }\\
\end{table}

\Leireftb{tb:exp:results} presents the quantitative results of different methods for torso reconstruction and electrode localization.
Here, \textit{SSM projection}-based electrode localization was first performed on an average torso, and then the coordinates were projected to each torso mesh based on the estimated transformation (see \Leireffig{fig:data:SSM}), acting as a baseline model for electrode localization \citep{conf/EMBC/smith2022}.  
\textit{PCN$_\text{encoder}$ + 3FC} referred to the electrode estimation directly from the extracted global features of PCN$_\text{encoder}$, followed by three fully connected (FC) layers.
In contrast, though \textit{TIM} also employed PCN$_\text{encoder}$ to extract both local and global point cloud features, it introduced the topology-informed keypoint detection component and keypoint-guided torso reconstruction module.
\textit{TIM \textit{w/o} kp} referred to the baseline PCN model \citep{conf/3DV/yuan2018}, without keypoint detection and surface skeleton assisted point cloud refinement module. 
\textit{TIM \textit{w/o} recon} still used keypoints as a guide for electrode localization, but the surface skeleton assisted torso reconstruction module was removed.

Conventional \textit{SSM projection}-based electrode localization involves iterative optimization, making it quite time-consuming (typically $30$-$35$ minutes). 
In contrast, deep learning-based algorithms, including both PCN and the proposed TIM methods, can estimate the electrode locations within seconds.
However, the accuracy of \textit{PCN$_\text{encoder}$ + 3FC} was inferior compared to both the \textit{SSM projection}-based localization and the proposed TIM methods.
Compared to \textit{PCN$_\text{encoder}$ + 3FC}, \textit{TIM \textit{w/o} kp} incorporated an additional torso reconstruction task, which significantly reduced the electrode localization error resulting from incomplete data ($1.54 \pm 0.446$ cm vs. $2.00 \pm 0.661$ cm). 
Conversely, \textit{TIM \textit{w/o} recon} focused on converting the electrode localization task into keypoint detection, demonstrating superior performance relative to solely introducing reconstruction. 
Our proposed method integrated both: it first performed keypoint detection and then created a surface skeleton from these keypoints to assist in torso reconstruction. 
It yielded the best results in electrode localization ($1.24 \pm 0.293$ cm). 
Moreover, the incorporation of the keypoint-based surface skeleton significantly improved torso reconstruction accuracy, achieving $2.38 \pm 0.306$ cm compared to $2.29 \pm 0.235$ cm (p-value = 0.004). 
We argue this might be because the surface skeleton module introduced topology attention for the torso reconstruction, especially on the torso area where electrodes are positioned.

\begin{figure*}[t]\center
 \includegraphics[width=0.86\textwidth]{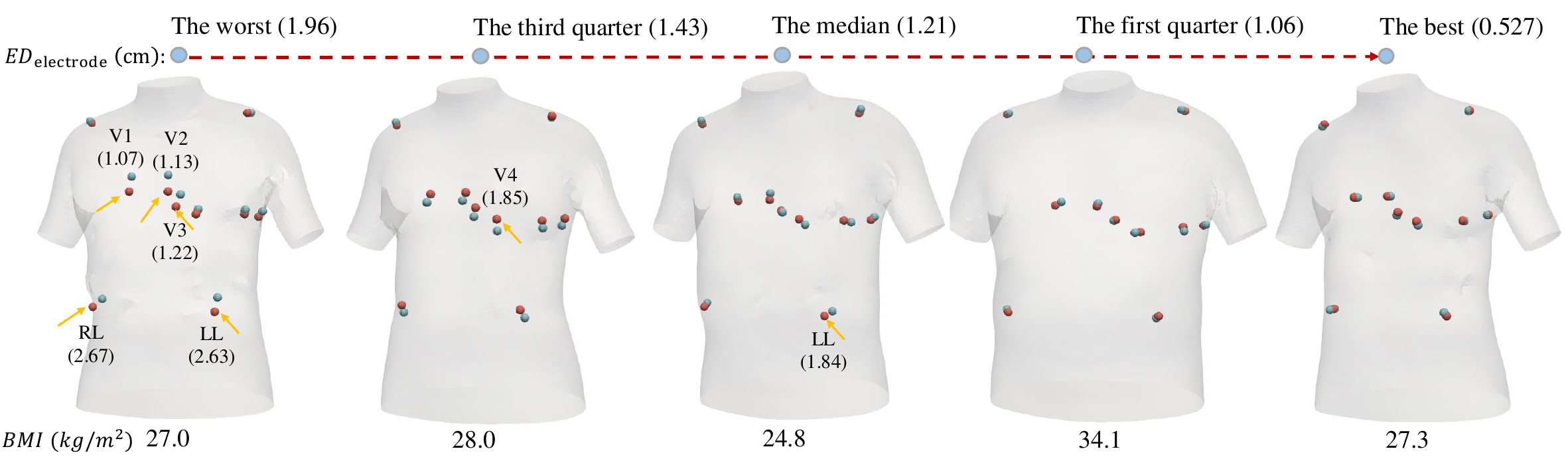}\\[-2ex] 
   \caption{Visualization of the electrode localization results by the proposed model at different levels of performance. The manual and automatic localized electrodes have been labeled in blue and red, respectively, overlapped on the ground truth torso. The average electrode localization error ($ED_\text{electrode}$) is shown at the top, while the body mass index (BMI) of each subject is presented at the bottom.} 
\label{fig:result:electrode visual}
\end{figure*}

\subsection{Results of the Proposed Method}

\subsubsection{Performance of the Proposed Method and Inter-Observer Study} 

The proposed method obtained a promising result for localizing electrodes, with an average ED of $ 1.24 \pm 0.293 $~cm.
There is no significant difference between males and females ($ 1.24 \pm 0.309 $~cm vs. $ 1.25 \pm 0.276 $~cm, p-value $=0.992$).
Note that, even for trained technicians, the electrode placement errors were often in the range of 2–3 cm and occasionally reached up to 6~cm \citep{journal/MBEC/kania2014}.
To better understand such variation in our dataset, we conducted an inter-observer study with two manual delineations.
Ten cases were randomly selected, and an additional expert manually localized electrodes. 
The inter-observer average ED was $ 1.31 \pm 0.373 $~cm, while our method achieved an average ED of $1.24 \pm 0.293$ cm, which is well within the range of expert variation.
This demonstrates that the accuracy of our method is comparable to the variability seen between different experts.


Figure~\ref{fig:result:electrode visual} provides 3D visualization of the electrodes overlapped on the ground truth torso geometry from five examples.
These five cases were the last, third quarter, median, first quarter, first cases from the test set in terms of $ED_\text{electrode}$ by the proposed method. 
This illustrates that the method could provide promising performance for localizing electrodes from incomplete torso contours.
In the worst case, we highlight the errors, particularly due to the shift of chest electrodes V1, V2, and V3 and limb electrodes LL and RL, pointed out by arrows.
In the third quarter and median cases, the errors were from the electrodes V4 and LL, respectively.
In general, the predicted electrodes LL and RL could be slightly higher, lower, or even outside the torso surface. 
Also, LL and RL exhibited similar shift directions as the chest electrodes.
In contrast, the electrode LA, RA, V5 and V6 can be accurately localized in all five cases.
More importantly, we observed significant shape variation among the five subjects, yet our method appears to handle this variation effectively.

\subsubsection{Performance of Different Electrodes} \label{result:different electrodes}

\begin{figure}[t]\center
 \includegraphics[width=0.5\textwidth]{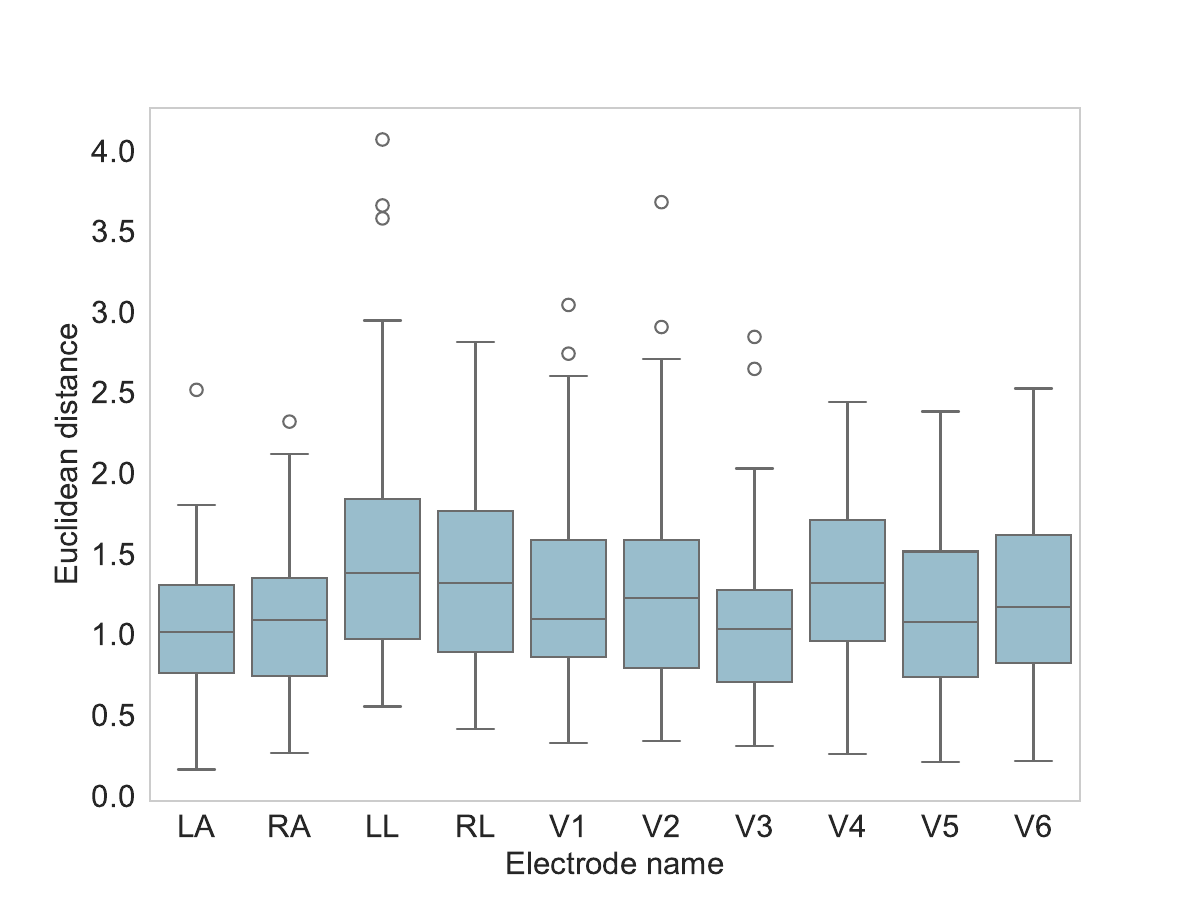}\\[-2ex] 
   \caption{Boxplots of the average Euclidean distance of electrode localization with respect to different electrode positions. } 
\label{fig:result:different_electrode}
\end{figure}

Figure~\ref{fig:result:different_electrode} presents the boxplots of the proposed algorithm for localizing different electrodes.
One can find that the performance varied among different electrodes. 
Specifically, limb electrodes LL and RL are generally more challenging to localize compared to chest electrodes (V1, V2, ..., V6), with localization errors of $1.49 \pm 0.587$~cm and $1.22 \pm 0.363$~cm, respectively. 
This is reasonable, as distant limb electrodes LL and RL are generally located in the peripheral areas of the torso, which are often outside the primary imaging region focused on the heart (especially electrode LL). 
Conversely, chest electrodes are positioned closer to the cardiac area, which is more accurately captured in imaging.
Variations in detecting electrodes LL and RL might also arise from their wider placement range without guidance from near landmarks.
Nonetheless, the displacements of distant limb electrodes are considered to have less impact on the ECG signal than shifts of chest electrodes \citep{journal/JCP/rajaganeshan2008,journal/MBEC/kania2014,journal/JE/zappon2024}, which has been further proved in our simulation study (see the Supplementary Material).
Among the chest electrodes, electrode V4 was the hardest to be accurately localized.
This may be because the torso contour in the region where V4 is placed can be less distinct or more variable among individuals, making it harder for the model to identify precise features for accurate localization compared to other electrodes with more prominent or consistent landmarks.

\begin{figure}[t]\center
 \includegraphics[width=0.5\textwidth]{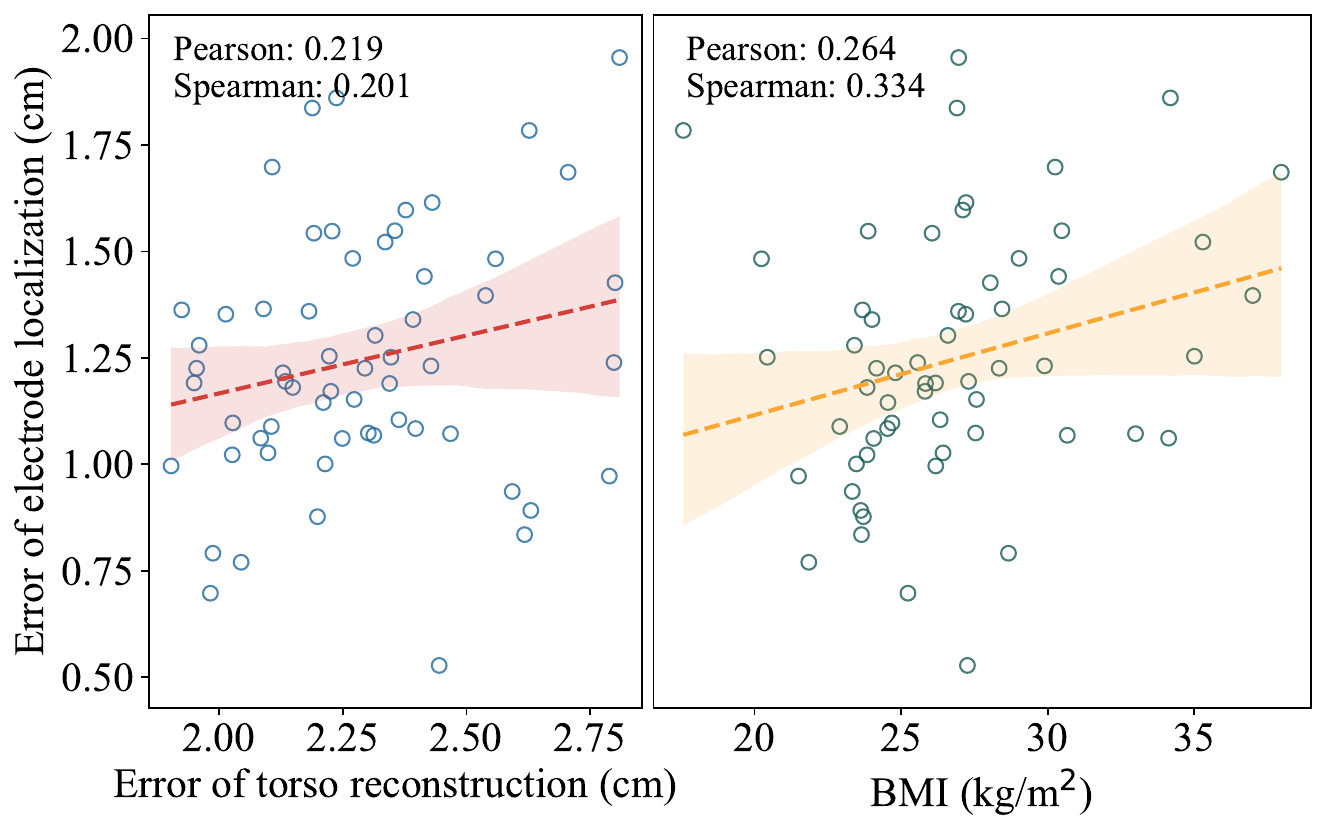}\\[-2ex]
   \caption{The scatter point plots and correlations between electrode localization error with respect to both torso reconstruction error and BMI.} 
\label{fig:result:correlation}
\end{figure}

\begin{figure*}[t]\center
 \includegraphics[width=0.125\textwidth]{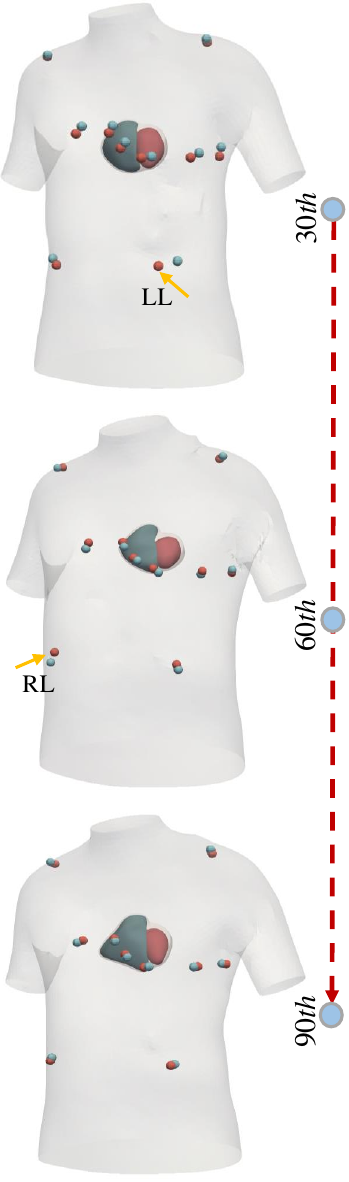}
 \includegraphics[width=0.78\textwidth]{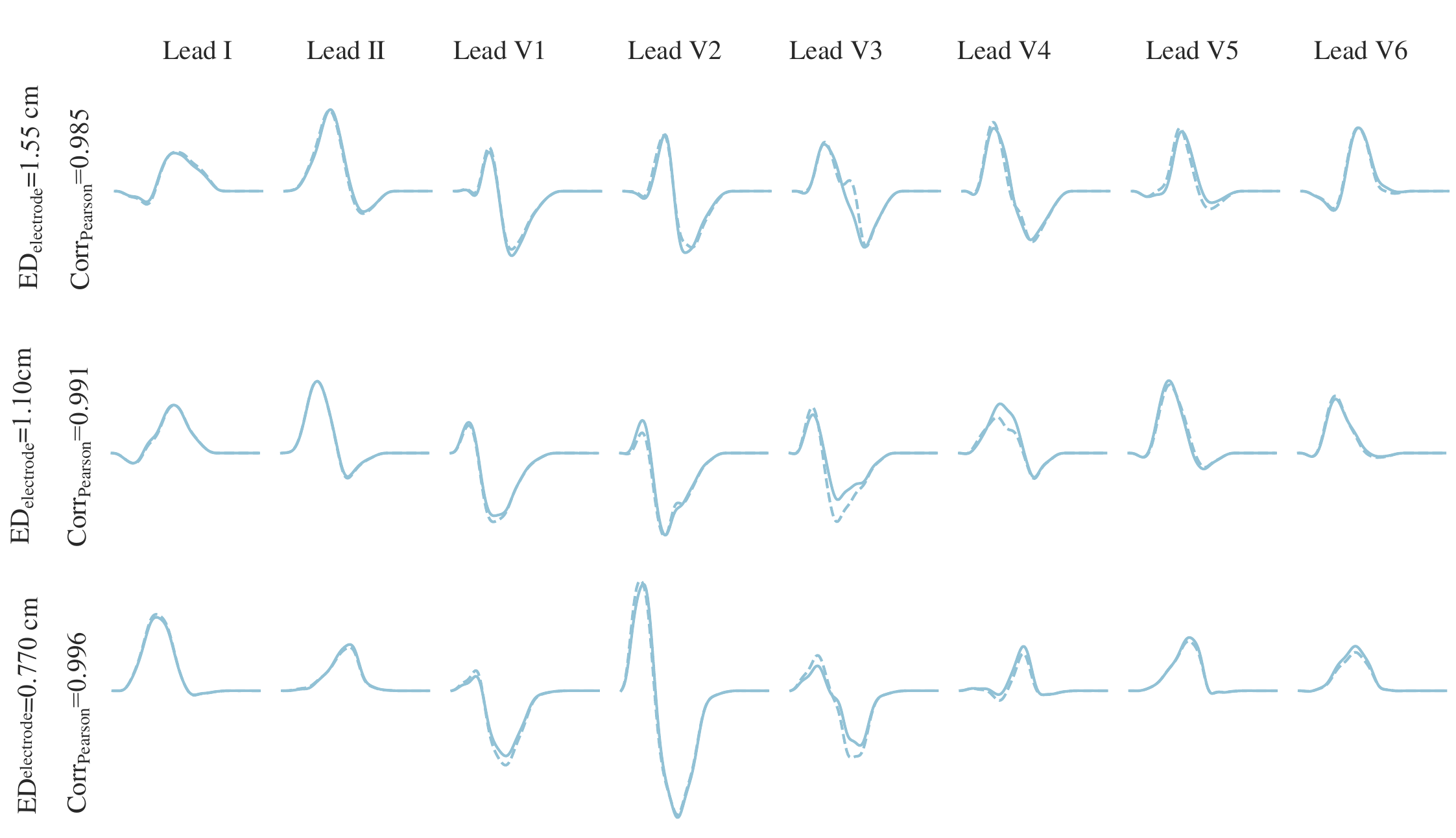}
\\[-2ex]
   \caption{Comparison of simulated ECG-QRS via predicted (in dashed lines) and ground truth electrodes (in solid lines). The left side shows the corresponding electrode locations, overlapped on the heart-torso anatomy. $Corr_{Pearson}$: Pearson correlation coefficient.} 
\label{fig:result:simulation}
\end{figure*}


\subsubsection{Effect of 3D Torso Reconstruction and BMI for Electrode Localization}

To analyze the relationship between the electrode localization error (\(ED_\text{electrode}\)) and the 3D torso reconstruction error (\(CD_\text{torso}\)) obtained by the proposed method, we plotted these values as two-dimensional scatter points for all test subjects, as shown in \Leireffig{fig:result:correlation}. 
The scatter plot indicates a positive but weak linear relationship between the two variables.
To quantify this observation, we performed linear regression, Pearson correlation, and Spearman’s rank correlation analyses.
The results of the linear regression analysis yielded an \( r^2 \) value of 0.048, suggesting that only 4.8\% of the variance in electrode localization error can be explained by the torso reconstruction error. 
This low \( r^2 \) value indicates a weak linear relationship between the two variables.
The Pearson and Spearman’s rank correlation coefficient was 0.219 and 0.201, respectively, further supporting the conclusion of a weak positive linear correlation between the two variables.
This implies that improvements in torso reconstruction accuracy do not necessarily lead to substantial reductions in electrode localization error. 
This may be attributed to the fact that electrodes (a subset of keypoints) can only partially represent the torso topology.

In addition, we examined the correlation between electrode localization error and BMI, given that the error reflects distance on the body. 
We observed a weak positive linear relationship, with Pearson and Spearman correlation coefficients of 0.264 and 0.334, respectively.
This outcome was expected, as higher BMI values typically correlate with greater body mass and potential changes in body geometry. 
However, our analysis found no significant correlation between torso reconstruction error and BMI (Pearson = 0.037, Spearman = 0.039), as presented in the Supplementary Material.
It indicates that while BMI influences localization accuracy, it does not impact torso reconstruction accuracy. 
Together, these results highlight the robustness of the proposed method in maintaining accuracy across a range of body geometries.

\subsection{\textit{In-Silico} ECG Simulation Evaluation} \label{exp:simulation}

\subsubsection{ECG Simulation Results}

To explore the feasibility of predicted electrodes for virtual heart modeling, we compared the simulated ECG-QRS based on predicted electrodes and ground truth electrodes.
Ten subjects were randomly selected from the test set for the \textit{in-silico} evaluation, and we automatically created their 3D biventricular heart models from cardiac MRIs based on the computational geometry based pipeline detailed in \citep{journal/PTRSA/banerjee2021}. 
Their torso reconstruction and electrode prediction error were $ 2.33 \pm 0.200 $ cm and $ 1.04 \pm 0.337 $ cm, respectively.
We employed an efficient orthotropic Eikonal model for the cardiac EP simulation with tuned parameters to ensure realistic ECG morphology \citep{journal/MedIA/camps2021,journal/MedIA/camps2024}.
Specifically, the body surface potential ($\Phi$) at the electrode position ($K_j$) was calculated as:
\begin{equation}
\label{eq:BSP}
\Phi(K_j) = \sum_{t=1}^{N_{src}} -\sigma_j (\nabla U)_t \left[\frac{\nabla b_t}{r_t}\right], 
\end{equation}
where $(\nabla U)_t$ represents the spatial gradient of membrane potential over the $t$-th tetrahedral element, $\sigma_j$ is the conductivity tensor for the $t$-th element, $b_j$ is the normalized volume scaling factor for the $t$-th element, $r_j$ is the Euclidean distance from the centroid of the $t$-th element to the electrode position ($K_j$), and $N_{src}$ is the total number of tetrahedral source elements. 
We employed the pseudo-ECG approach to model the heart in an infinite, isotropic, homogeneously conductive medium, enabling unipolar extracellular potential computation via an integral formulation, as presented in \Leirefeq{eq:BSP}. 
These unipolar electrograms were then used to derive the three Einthoven limb leads and the Wilson precordial leads.
Compared to finite volume conductor methods \citep{conf/FIMH/ogiermann2021} and heterogeneous conductivity models \citep{journal/TBME/bishop2011}, the primary differences observed are in amplitude, with only minor variations in QRS morphology \citep{journal/TBME/keller2010}.
To assess QRS morphology, we used the dynamic time warping (DTW) distance \citep{conf/CSMC/tuzcu2005,conf/FIMH/li2023} as a ECG misalignment measurement, yielding an average DTW distance of $0.081 \pm 0.033$.
Furthermore, we calculated their Pearson correlation coefficient ($Corr_{Pearson}$) as a measurement of the similarity between two ECG signals, with average correlation coefficient of $0.989 \pm 0.013$.

\Leireffigure{fig:result:simulation} presents the obtained simulated QRS morphology at different leads using predicted and ground truth electrodes from three subjects.
Note that here a low-pass filter was applied to reduce ECG noise artifacts from coarse mesh simulation, with signals aligned to start at 0 standardized voltage units.
We only presented eight independent leads, as lead III and the augmented leads (aVF, aVR, and aVL) are linear combinations of other leads.
These three cases were the 30th, 60th, and 90th percentiles from the 10 cases in terms of $ED_\text{electrode}$ by the proposed method. 
One can see that the simulated ECG from the automatically predicted electrodes closely aligned with that from the ground truth electrodes.
No significant changes in ECG morphology were observed in leads I, II, and V6, with average $Corr_{Pearson}$ values exceeding $0.998$.
Only slight misalignments were observed in certain leads, notably V3 and V4, with average $Corr_{Pearson}$ values of $0.962 \pm 0.069$ and $0.977 \pm 0.020$, respectively.
These are generally consistent with the conclusion reported in literature \citep{journal/MBEC/kania2014,journal/JE/roudijk2021}.
Additionally, we observed that their QRS durations were consistent across all subjects, while their average R/S amplitude ratio difference was $0.058 \pm 0.035$.
As expected, a decrease in electrode localization errors generally leads to a reduction in observed misalignment.

\begin{figure}[t]\center
 \includegraphics[width=0.5\textwidth]{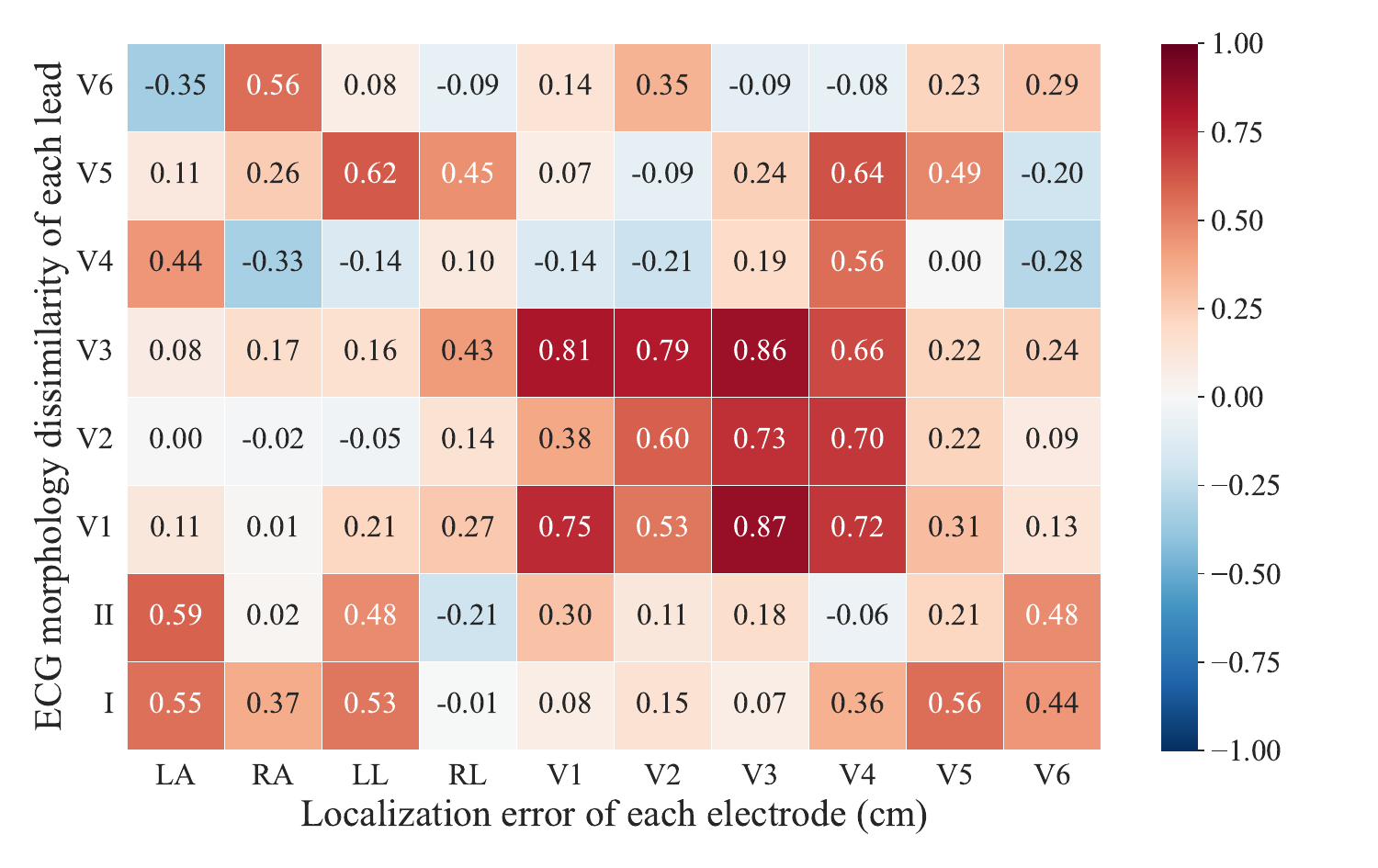} \\[-2ex]
   \caption{Correlation heatmap between the localization error of 12-lead ECG electrodes and the morphology misalignment of different leads.} 
\label{fig:result:correlation map}
\end{figure}

\subsubsection{Correlation between ECG Electrodes and Leads}

To quantitatively assess the relationship between electrode localization error and ECG morphology misalignment (DTW), we conducted a linear regression analysis. 
The resulting linear coefficient $r^2$ was 0.576, indicating a moderately strong positive linear correlation between the localization error and morphology misalignment.
To further analyze the relationship between different electrodes and leads, we calculated their Pearson correlation for the ten subjects.
\Leireffigure{fig:result:correlation map} presents the correlation heatmap between different electrodes and ECG leads.
We observe a strong positive correlation between electrodes V1-V5 and their corresponding ECG leads V1-V5, with an especially high correlation for lead V3. 
This indicates that misplacement of electrodes in the chest area will significantly impact the morphology of their associated precordial leads, which is consistent with observations from our uncertainty study (see Supplementary Material). 
Additionally, electrodes LA, RA, and LL show a notable correlation with leads I and II, aligning with the findings of the uncertainty study. 
However, there are some unexpected correlations between certain leads and electrodes, which may indicate complex interactions due to individual anatomical differences or variations in electrode placement.
Note that this analysis is based on a limited sample size of 10 subjects, which may introduce potential biases.


\section{Discussion and Conclusion}

In this study, we have presented a fully automatic framework for simultaneous 12-lead ECG standard torso electrode localization and 3D torso geometry reconstruction from cardiac MRI data. 
The proposed model was evaluated on 200 subjects from the UK Biobank, yielding better results than the conventional SSM project-based method ($1.24 \pm 0.293$ cm vs. $1.48 \pm 0.362$ cm) and significantly improved efficiency ($2$ s vs. $30$-$35$~min).
This could attribute to both the topology-guided keypoint detection and the surface skeleton-assisted PCN modules, as we demonstrated in Sec.~\ref{exp:compare}.
Also, we found that random resampling of input contours had a significant impact on the performance, which might be attributed to the sparsity and incompleteness of the contours. 
However, this issue can be effectively addressed by increasing the number of random resampling iterations for data augmentation (see Sec. \ref{exp:parameter study:rr}).
The number of keypoints also affected model performance, namely they need to adequately cover the electrodes without being excessively numerous.
We found that localizing electrodes LL and RL accurately was challenging, likely due to the lack of guidance from nearby landmarks (see Sec.~\ref{result:different electrodes}).
Notably, the simulated ECG morphology based on predicted electrodes exhibited good alignment with that based on ground truth (see Sec. \ref{exp:simulation}), affirming the feasibility of employing clinically standard cardiac MRI for efficient and personalized virtual heart modeling. 

Nonetheless, there are three limitations of this work. 
Firstly, the proposed method aims to estimate the 12-lead ECG electrode position based on the standard protocol, assuming no significant placement errors. According to the literature, typical electrode placement errors are often within the range of 2–3 cm \citep{journal/MBEC/kania2014}. 
However, this method may not be applicable if the actual electrode placement deviates substantially from the standard protocol.
Secondly, there are few related studies in the literature, and no state-of-the-art 12-lead ECG electrode localization results available for comparison besides \citet{journal/JE/roudijk2021}. 
They employed 3D camera to localize 12-lead ECG electrodes for 20 atrial fibrillation patients, to reduce electrode repositioning errors during longitudinal ECG acquisition.
They reported a median distance of 1.00 [0.64–1.52] cm between 3D guided repositioned electrodes, which was similar to our results ($1.24 \pm 0.293$ cm).
We further evaluated the performance of the proposed method for \textit{in-silico} ECG simulation, providing a benchmark and demonstrating its potential effectiveness in enabling an efficient CDT.
Thirdly, the method has only been tested on a limited dataset for \textit{in-silico} ECG simulation, and its generalizability to other datasets and populations remains to be validated.
Also, this \textit{in-silico} evaluation focused exclusively on the QRS complex rather than the whole ECG signal. 
As reported in \citet{journal/JE/zappon2024}, higher placement of electrodes V1 and V2 can cause noticeable T-wave inversion in ECG leads V1 and V2.

In our future work, we plan to detect the actual electrode positions by adopting the dataset collection approach described by \citet{journal/MedIA/gillette2021}, which employed MRI-compatible electrodes positioned before image acquisition.
Furthermore, we will explore a more detailed patient-specific heart-torso anatomy, incorporating a whole heart embedded within a torso, with distinct volumes for cardiac blood pools, lungs, bones, liver, fat, and skin, to enhance the fidelity of the CDT. 
We will investigate the heterogeneous conductivity within the torso which presents substantial differences between compartments (including organs or distinct tissues such as bones, fat, and skin).
At the same time, the electrode localization can be more accurately localized on the finer torso geometry with bone structures.
We can evaluate the proposed model on a larger population dataset from different centers, to investigate its generalization ability.
On this larger dataset, we can perform a comprehensive sensitivity analysis to qualitatively and quantitatively investigate the impact of the uncertainties from torso and electrodes on the ECG morphology.
These uncertainties, along with potential device artifacts in clinical ECGs, must be considered when applying predicted electrode positions to CDTs. 
Such factors can introduce variability in ECG morphology, potentially affecting the parameterization and accuracy of CDTs.
Also, we will extend the \textit{in-silico} ECG simulation to the entire ECG signal.
We will then perform inverse inference to calibrate the simulated ECG, ensuring it aligns accurately with clinically measured ECG data. 
To achieve this, we will explore a computationally efficient surrogate to replace current simulation model, to further speed the creation of CDT.
In conclusion, this work demonstrates the feasibility of automatic electrode detection using clinically standard cardiac MRI, paving the way for future research to explore real-time applications of CDTs.


\section*{Acknowledgments}
This work was supported by start-up funding from Soton and NUS to L. Li and
CompBiomedX EPSRC-funded grant (EP/X019446/1) and a Wellcome Trust Senior Research Fellowship in Basic Biomedical Sciences (214290/Z/18/Z) to B. Rodriguez.
H. Smith was funded by Wellcome Trust Studentship under Grant No. 102161/Z/13/Z. 
A. Banerjee was supported by the Royal Society University Research Fellowship (Grant No. URF{\textbackslash}R1{\textbackslash}221314).

\bibliographystyle{model2-names}
\biboptions{authoryear}
\bibliography{A_refs}

\begin{thebibliography}{51}
\expandafter\ifx\csname natexlab\endcsname\relax\def\natexlab#1{#1}\fi
\providecommand{\url}[1]{\texttt{#1}}
\providecommand{\href}[2]{#2}
\providecommand{\path}[1]{#1}
\providecommand{\DOIprefix}{doi:}
\providecommand{\ArXivprefix}{arXiv:}
\providecommand{\URLprefix}{URL: }
\providecommand{\Pubmedprefix}{pmid:}
\providecommand{\doi}[1]{\href{http://dx.doi.org/#1}{\path{#1}}}
\providecommand{\Pubmed}[1]{\href{pmid:#1}{\path{#1}}}
\providecommand{\bibinfo}[2]{#2}
\ifx\xfnm\relax \def\xfnm[#1]{\unskip,\space#1}\fi
\bibitem[{Arevalo et~al.(2016)Arevalo, Vadakkumpadan, Guallar, Jebb, Malamas, Wu and Trayanova}]{journal/Nature_Comm/arevalo2016}
\bibinfo{author}{Arevalo, H.J.}, \bibinfo{author}{Vadakkumpadan, F.}, \bibinfo{author}{Guallar, E.}, \bibinfo{author}{Jebb, A.}, \bibinfo{author}{Malamas, P.}, \bibinfo{author}{Wu, K.C.}, \bibinfo{author}{Trayanova, N.}, \bibinfo{year}{2016}.
\newblock \bibinfo{title}{Arrhythmia risk stratification of patients after myocardial infarction using personalized heart models}.
\newblock \bibinfo{journal}{Nature Communications} \bibinfo{volume}{7}, \bibinfo{pages}{11437}.
\bibitem[{Aste et~al.(2015)Aste, Boninsegna, Freno and Trentin}]{journal/PAA/aste2015}
\bibinfo{author}{Aste, M.}, \bibinfo{author}{Boninsegna, M.}, \bibinfo{author}{Freno, A.}, \bibinfo{author}{Trentin, E.}, \bibinfo{year}{2015}.
\newblock \bibinfo{title}{Techniques for dealing with incomplete data: a tutorial and survey}.
\newblock \bibinfo{journal}{Pattern Analysis and Applications} \bibinfo{volume}{18}, \bibinfo{pages}{1--29}.
\bibitem[{Banerjee et~al.(2021)Banerjee, Camps, Zacur, Andrews, Rudy, Choudhury, Rodriguez and Grau}]{journal/PTRSA/banerjee2021}
\bibinfo{author}{Banerjee, A.}, \bibinfo{author}{Camps, J.}, \bibinfo{author}{Zacur, E.}, \bibinfo{author}{Andrews, C.M.}, \bibinfo{author}{Rudy, Y.}, \bibinfo{author}{Choudhury, R.P.}, \bibinfo{author}{Rodriguez, B.}, \bibinfo{author}{Grau, V.}, \bibinfo{year}{2021}.
\newblock \bibinfo{title}{A completely automated pipeline for {3D} reconstruction of human heart from {2D} cine magnetic resonance slices}.
\newblock \bibinfo{journal}{Philosophical Transactions of the Royal Society A} \bibinfo{volume}{379}, \bibinfo{pages}{20200257}.
\bibitem[{Beetz et~al.(2023)Beetz, Banerjee, Ossenberg-Engels and Grau}]{journal/MedIA/beetz2023}
\bibinfo{author}{Beetz, M.}, \bibinfo{author}{Banerjee, A.}, \bibinfo{author}{Ossenberg-Engels, J.}, \bibinfo{author}{Grau, V.}, \bibinfo{year}{2023}.
\newblock \bibinfo{title}{Multi-class point cloud completion networks for {3D} cardiac anatomy reconstruction from cine magnetic resonance images}.
\newblock \bibinfo{journal}{Medical Image Analysis} \bibinfo{volume}{90}, \bibinfo{pages}{102975}.
\bibitem[{Bishop and Plank(2011)}]{journal/TBME/bishop2011}
\bibinfo{author}{Bishop, M.J.}, \bibinfo{author}{Plank, G.}, \bibinfo{year}{2011}.
\newblock \bibinfo{title}{Bidomain ecg simulations using an augmented monodomain model for the cardiac source}.
\newblock \bibinfo{journal}{IEEE transactions on biomedical engineering} \bibinfo{volume}{58}, \bibinfo{pages}{2297--2307}.
\bibitem[{Boyle et~al.(2019)Boyle, Zghaib, Zahid, Ali, Deng, Franceschi, Hakim, Murphy, Prakosa, Zimmerman, Ashikaga, Marine, Kolandaivelu, Nazarian, Spragg, Calkins and Trayanova}]{journal/Nature_BME/boyle2019}
\bibinfo{author}{Boyle, P.M.}, \bibinfo{author}{Zghaib, T.}, \bibinfo{author}{Zahid, S.}, \bibinfo{author}{Ali, R.L.}, \bibinfo{author}{Deng, D.}, \bibinfo{author}{Franceschi, W.H.}, \bibinfo{author}{Hakim, J.B.}, \bibinfo{author}{Murphy, M.J.}, \bibinfo{author}{Prakosa, A.}, \bibinfo{author}{Zimmerman, S.L.}, \bibinfo{author}{Ashikaga, H.}, \bibinfo{author}{Marine, J.E.}, \bibinfo{author}{Kolandaivelu, A.}, \bibinfo{author}{Nazarian, S.}, \bibinfo{author}{Spragg, D.D.}, \bibinfo{author}{Calkins, H.}, \bibinfo{author}{Trayanova, N.A.}, \bibinfo{year}{2019}.
\newblock \bibinfo{title}{Computationally guided personalized targeted ablation of persistent atrial fibrillation}.
\newblock \bibinfo{journal}{Nature Biomedical Engineering} \bibinfo{volume}{3}, \bibinfo{pages}{870--879}.
\bibitem[{Camps et~al.(2024a)Camps, Berg, Wang, Sebastian, Riebel, Doste, Zhou, Sachetto, Coleman, Lawson, Grau, Burrage, Bueno-Orovio, Weber~dos Santos and Rodriguez}]{journal/MedIA/camps2024a}
\bibinfo{author}{Camps, J.}, \bibinfo{author}{Berg, L.A.}, \bibinfo{author}{Wang, Z.J.}, \bibinfo{author}{Sebastian, R.}, \bibinfo{author}{Riebel, L.L.}, \bibinfo{author}{Doste, R.}, \bibinfo{author}{Zhou, X.}, \bibinfo{author}{Sachetto, R.}, \bibinfo{author}{Coleman, J.}, \bibinfo{author}{Lawson, B.}, \bibinfo{author}{Grau, V.}, \bibinfo{author}{Burrage, K.}, \bibinfo{author}{Bueno-Orovio, A.}, \bibinfo{author}{Weber~dos Santos, R.}, \bibinfo{author}{Rodriguez, B.}, \bibinfo{year}{2024}a.
\newblock \bibinfo{title}{Digital twinning of the human ventricular activation sequence to clinical 12-lead {ECG}s and magnetic resonance imaging using realistic purkinje networks for in silico clinical trials}.
\newblock \bibinfo{journal}{Medical Image Analysis} \bibinfo{volume}{94}, \bibinfo{pages}{103108}.
\bibitem[{Camps et~al.(2021)Camps, Lawson, Drovandi, Minchole, Wang, Grau, Burrage and Rodriguez}]{journal/MedIA/camps2021}
\bibinfo{author}{Camps, J.}, \bibinfo{author}{Lawson, B.}, \bibinfo{author}{Drovandi, C.}, \bibinfo{author}{Minchole, A.}, \bibinfo{author}{Wang, Z.J.}, \bibinfo{author}{Grau, V.}, \bibinfo{author}{Burrage, K.}, \bibinfo{author}{Rodriguez, B.}, \bibinfo{year}{2021}.
\newblock \bibinfo{title}{Inference of ventricular activation properties from non-invasive electrocardiography}.
\newblock \bibinfo{journal}{Medical Image Analysis} \bibinfo{volume}{73}, \bibinfo{pages}{102143}.
\bibitem[{Camps et~al.(2024b)Camps, Wang, Doste, Berg, Holmes, Lawson, Tomek, Burrage, Bueno-Orovio and Rodriguez}]{journal/MedIA/camps2024}
\bibinfo{author}{Camps, J.}, \bibinfo{author}{Wang, Z.J.}, \bibinfo{author}{Doste, R.}, \bibinfo{author}{Berg, L.A.}, \bibinfo{author}{Holmes, M.}, \bibinfo{author}{Lawson, B.}, \bibinfo{author}{Tomek, J.}, \bibinfo{author}{Burrage, K.}, \bibinfo{author}{Bueno-Orovio, A.}, \bibinfo{author}{Rodriguez, B.}, \bibinfo{year}{2024}b.
\newblock \bibinfo{title}{Harnessing 12-lead {ECG} and {MRI} data to personalise repolarisation profiles in cardiac digital twin models for enhanced virtual drug testing}.
\newblock \bibinfo{journal}{Medical Image Analysis} , \bibinfo{pages}{103361}.
\bibitem[{van Dam et~al.(2014)van Dam, Gordon and Laks}]{journal/JE/van2014}
\bibinfo{author}{van Dam, P.M.}, \bibinfo{author}{Gordon, J.P.}, \bibinfo{author}{Laks, M.}, \bibinfo{year}{2014}.
\newblock \bibinfo{title}{Sensitivity of {CIPS}-computed {PVC} location to measurement errors in {ECG} electrode position: the need for the {3D} camera}.
\newblock \bibinfo{journal}{Journal of Electrocardiology} \bibinfo{volume}{47}, \bibinfo{pages}{788--793}.
\bibitem[{Geneser et~al.(2008)Geneser, Kirby and MacLeod}]{journal/TBME/ghanem2003}
\bibinfo{author}{Geneser, S.E.}, \bibinfo{author}{Kirby, R.M.}, \bibinfo{author}{MacLeod, R.S.}, \bibinfo{year}{2008}.
\newblock \bibinfo{title}{Application of stochastic finite element methods to study the sensitivity of ecg forward modeling to organ conductivity}.
\newblock \bibinfo{journal}{IEEE Transactions on Biomedical Engineering} \bibinfo{volume}{55}, \bibinfo{pages}{31--40}.
\bibitem[{Ghanem et~al.(2003)Ghanem, Ramanathan, Jia and Rudy}]{journal/TMI/ghanem2003}
\bibinfo{author}{Ghanem, R.N.}, \bibinfo{author}{Ramanathan, C.}, \bibinfo{author}{Jia, P.}, \bibinfo{author}{Rudy, Y.}, \bibinfo{year}{2003}.
\newblock \bibinfo{title}{Heart-surface reconstruction and {ECG} electrodes localization using fluoroscopy, epipolar geometry and stereovision: application to noninvasive imaging of cardiac electrical activity}.
\newblock \bibinfo{journal}{IEEE Transactions on Medical Imaging} \bibinfo{volume}{22}, \bibinfo{pages}{1307--1318}.
\bibitem[{Giffard-Roisin et~al.(2016)Giffard-Roisin, Jackson, Fovargue, Lee, Delingette, Razavi, Ayache and Sermesant}]{journal/TBME/giffard2016}
\bibinfo{author}{Giffard-Roisin, S.}, \bibinfo{author}{Jackson, T.}, \bibinfo{author}{Fovargue, L.}, \bibinfo{author}{Lee, J.}, \bibinfo{author}{Delingette, H.}, \bibinfo{author}{Razavi, R.}, \bibinfo{author}{Ayache, N.}, \bibinfo{author}{Sermesant, M.}, \bibinfo{year}{2016}.
\newblock \bibinfo{title}{Noninvasive personalization of a cardiac electrophysiology model from body surface potential mapping}.
\newblock \bibinfo{journal}{IEEE Transactions on Biomedical Engineering} \bibinfo{volume}{64}, \bibinfo{pages}{2206--2218}.
\bibitem[{Gillette et~al.(2021)Gillette, Gsell, Prassl, Karabelas, Reiter, Reiter, Grandits, Payer, {\v{S}}tern, Urschler, Bayer, Augustin, Neic, Pock, Vigmond and Plank}]{journal/MedIA/gillette2021}
\bibinfo{author}{Gillette, K.}, \bibinfo{author}{Gsell, M.A.}, \bibinfo{author}{Prassl, A.J.}, \bibinfo{author}{Karabelas, E.}, \bibinfo{author}{Reiter, U.}, \bibinfo{author}{Reiter, G.}, \bibinfo{author}{Grandits, T.}, \bibinfo{author}{Payer, C.}, \bibinfo{author}{{\v{S}}tern, D.}, \bibinfo{author}{Urschler, M.}, \bibinfo{author}{Bayer, J.D.}, \bibinfo{author}{Augustin, C.M.}, \bibinfo{author}{Neic, A.}, \bibinfo{author}{Pock, T.}, \bibinfo{author}{Vigmond, E.J.}, \bibinfo{author}{Plank, G.}, \bibinfo{year}{2021}.
\newblock \bibinfo{title}{A framework for the generation of digital twins of cardiac electrophysiology from clinical 12-leads {ECG}s}.
\newblock \bibinfo{journal}{Medical Image Analysis} \bibinfo{volume}{71}, \bibinfo{pages}{102080}.
\bibitem[{Gillette et~al.(2022)Gillette, Gsell, Strocchi, Grandits, Neic, Manninger, Scherr, Roney, Prassl, Augustin, Vigmond and Plank}]{journal/FiP/gillette2022}
\bibinfo{author}{Gillette, K.}, \bibinfo{author}{Gsell, M.A.}, \bibinfo{author}{Strocchi, M.}, \bibinfo{author}{Grandits, T.}, \bibinfo{author}{Neic, A.}, \bibinfo{author}{Manninger, M.}, \bibinfo{author}{Scherr, D.}, \bibinfo{author}{Roney, C.H.}, \bibinfo{author}{Prassl, A.J.}, \bibinfo{author}{Augustin, C.M.}, \bibinfo{author}{Vigmond, E.J.}, \bibinfo{author}{Plank, G.}, \bibinfo{year}{2022}.
\newblock \bibinfo{title}{A personalized real-time virtual model of whole heart electrophysiology}.
\newblock \bibinfo{journal}{Frontiers in Physiology} , \bibinfo{pages}{1860}.
\bibitem[{Gillette et~al.(2015)Gillette, Tate, Kindall, Van~Dam, Kholmovski and MacLeod}]{conf/CiC/gillette2015}
\bibinfo{author}{Gillette, K.}, \bibinfo{author}{Tate, J.}, \bibinfo{author}{Kindall, B.}, \bibinfo{author}{Van~Dam, P.}, \bibinfo{author}{Kholmovski, E.}, \bibinfo{author}{MacLeod, R.S.}, \bibinfo{year}{2015}.
\newblock \bibinfo{title}{Generation of combined-modality tetrahedral meshes}, in: \bibinfo{booktitle}{Computing in Cardiology Conference}, \bibinfo{organization}{IEEE}. pp. \bibinfo{pages}{953--956}.
\bibitem[{Kania et~al.(2014)Kania, Rix, Fereniec, Zavala-Fernandez, Janusek, Mroczka, Stix and Maniewski}]{journal/MBEC/kania2014}
\bibinfo{author}{Kania, M.}, \bibinfo{author}{Rix, H.}, \bibinfo{author}{Fereniec, M.}, \bibinfo{author}{Zavala-Fernandez, H.}, \bibinfo{author}{Janusek, D.}, \bibinfo{author}{Mroczka, T.}, \bibinfo{author}{Stix, G.}, \bibinfo{author}{Maniewski, R.}, \bibinfo{year}{2014}.
\newblock \bibinfo{title}{The effect of precordial lead displacement on {ECG} morphology}.
\newblock \bibinfo{journal}{Medical \& Biological Engineering \& Computing} \bibinfo{volume}{52}, \bibinfo{pages}{109--119}.
\bibitem[{Keller et~al.(2010)Keller, Weber, Seemann and D{\"o}ssel}]{journal/TBME/keller2010}
\bibinfo{author}{Keller, D.U.}, \bibinfo{author}{Weber, F.M.}, \bibinfo{author}{Seemann, G.}, \bibinfo{author}{D{\"o}ssel, O.}, \bibinfo{year}{2010}.
\newblock \bibinfo{title}{Ranking the influence of tissue conductivities on forward-calculated {ECG}s}.
\newblock \bibinfo{journal}{IEEE Transactions on Biomedical Engineering} \bibinfo{volume}{57}, \bibinfo{pages}{1568--1576}.
\bibitem[{Kramer et~al.(2020)Kramer, Barkhausen, Bucciarelli-Ducci, Flamm, Kim and Nagel}]{journal/JCMR/kramer2020}
\bibinfo{author}{Kramer, C.M.}, \bibinfo{author}{Barkhausen, J.}, \bibinfo{author}{Bucciarelli-Ducci, C.}, \bibinfo{author}{Flamm, S.D.}, \bibinfo{author}{Kim, R.J.}, \bibinfo{author}{Nagel, E.}, \bibinfo{year}{2020}.
\newblock \bibinfo{title}{Standardized cardiovascular magnetic resonance imaging (cmr) protocols: 2020 update}.
\newblock \bibinfo{journal}{Journal of Cardiovascular Magnetic Resonance} \bibinfo{volume}{22}, \bibinfo{pages}{17}.
\bibitem[{Li et~al.(2024a)Li, Camps, Rodriguez and Grau}]{journal/RBME/li2024}
\bibinfo{author}{Li, L.}, \bibinfo{author}{Camps, J.}, \bibinfo{author}{Rodriguez, B.}, \bibinfo{author}{Grau, V.}, \bibinfo{year}{2024}a.
\newblock \bibinfo{title}{Solving the inverse problem of electrocardiography for cardiac digital twins: A survey}.
\newblock \bibinfo{journal}{IEEE Reviews in Biomedical Engineering} .
\bibitem[{Li et~al.(2023a)Li, Camps, Wang, Banerjee, Rodriguez and Grau}]{conf/FIMH/li2023}
\bibinfo{author}{Li, L.}, \bibinfo{author}{Camps, J.}, \bibinfo{author}{Wang, Z.}, \bibinfo{author}{Banerjee, A.}, \bibinfo{author}{Rodriguez, B.}, \bibinfo{author}{Grau, V.}, \bibinfo{year}{2023}a.
\newblock \bibinfo{title}{Influence of myocardial infarction on {QRS} properties: A simulation study}, in: \bibinfo{booktitle}{International Conference on Functional Imaging and Modeling of the Heart}, \bibinfo{organization}{Springer}. pp. \bibinfo{pages}{223--232}.
\bibitem[{Li et~al.(2024b)Li, Camps, Wang, Beetz, Banerjee, Rodriguez and Grau}]{journal/TMI/li2024}
\bibinfo{author}{Li, L.}, \bibinfo{author}{Camps, J.}, \bibinfo{author}{Wang, Z.}, \bibinfo{author}{Beetz, M.}, \bibinfo{author}{Banerjee, A.}, \bibinfo{author}{Rodriguez, B.}, \bibinfo{author}{Grau, V.}, \bibinfo{year}{2024}b.
\newblock \bibinfo{title}{Towards enabling cardiac digital twins of myocardial infarction using deep computational models for inverse inference}.
\newblock \bibinfo{journal}{IEEE Transactions on Medical Imaging} \bibinfo{volume}{43}, \bibinfo{pages}{2466 -- 2478}.
\bibitem[{Li et~al.(2023b)Li, Ding, Huang, Zhuang and Grau}]{journal/MedIA/li2023}
\bibinfo{author}{Li, L.}, \bibinfo{author}{Ding, W.}, \bibinfo{author}{Huang, L.}, \bibinfo{author}{Zhuang, X.}, \bibinfo{author}{Grau, V.}, \bibinfo{year}{2023}b.
\newblock \bibinfo{title}{Multi-modality cardiac image computing: A survey}.
\newblock \bibinfo{journal}{Medical Image Analysis} , \bibinfo{pages}{102869}.
\bibitem[{Loewe et~al.(2022)Loewe, Mart{\'\i}nez~D{\'\i}az, Nagel and S{\'a}nchez}]{book/ITSCE/loewe2022}
\bibinfo{author}{Loewe, A.}, \bibinfo{author}{Mart{\'\i}nez~D{\'\i}az, P.}, \bibinfo{author}{Nagel, C.}, \bibinfo{author}{S{\'a}nchez, J.}, \bibinfo{year}{2022}.
\newblock \bibinfo{title}{Cardiac digital twin modeling}, in: \bibinfo{booktitle}{Innovative treatment strategies for clinical electrophysiology}. \bibinfo{publisher}{Springer}, pp. \bibinfo{pages}{111--134}.
\bibitem[{Man et~al.(2008)Man, Maan, Kim, Draisma, Schalij, van~der Wall and Swenne}]{journal/JE/man2008}
\bibinfo{author}{Man, S.C.}, \bibinfo{author}{Maan, A.C.}, \bibinfo{author}{Kim, E.}, \bibinfo{author}{Draisma, H.H.}, \bibinfo{author}{Schalij, M.J.}, \bibinfo{author}{van~der Wall, E.E.}, \bibinfo{author}{Swenne, C.A.}, \bibinfo{year}{2008}.
\newblock \bibinfo{title}{Reconstruction of standard 12-lead electrocardiograms from 12-lead electrocardiograms recorded with the {M}ason-{L}ikar electrode configuration}.
\newblock \bibinfo{journal}{Journal of Electrocardiology} \bibinfo{volume}{41}, \bibinfo{pages}{211--219}.
\bibitem[{Minchol{\'e} et~al.(2019)Minchol{\'e}, Zacur, Ariga, Grau and Rodriguez}]{journal/FiP/minchole2019}
\bibinfo{author}{Minchol{\'e}, A.}, \bibinfo{author}{Zacur, E.}, \bibinfo{author}{Ariga, R.}, \bibinfo{author}{Grau, V.}, \bibinfo{author}{Rodriguez, B.}, \bibinfo{year}{2019}.
\newblock \bibinfo{title}{{MRI}-based computational torso/biventricular multiscale models to investigate the impact of anatomical variability on the {ECG QRS} complex}.
\newblock \bibinfo{journal}{Frontiers in Physiology} \bibinfo{volume}{10}, \bibinfo{pages}{1103}.
\bibitem[{Ogiermann et~al.(2021)Ogiermann, Balzani and Perotti}]{conf/FIMH/ogiermann2021}
\bibinfo{author}{Ogiermann, D.}, \bibinfo{author}{Balzani, D.}, \bibinfo{author}{Perotti, L.E.}, \bibinfo{year}{2021}.
\newblock \bibinfo{title}{The effect of modeling assumptions on the {ECG} in monodomain and bidomain simulations}, in: \bibinfo{booktitle}{International Conference on Functional Imaging and Modeling of the Heart}, \bibinfo{organization}{Springer}. pp. \bibinfo{pages}{503--514}.
\bibitem[{Perez-Alday et~al.(2018)Perez-Alday, Thomas, Kabir, Sedaghat, Rogovoy, van Dam, van Dam, Woodward, Fuss, Ferencik and Tereshchenko}]{journal/JE/perez2018}
\bibinfo{author}{Perez-Alday, E.A.}, \bibinfo{author}{Thomas, J.A.}, \bibinfo{author}{Kabir, M.}, \bibinfo{author}{Sedaghat, G.}, \bibinfo{author}{Rogovoy, N.}, \bibinfo{author}{van Dam, E.}, \bibinfo{author}{van Dam, P.}, \bibinfo{author}{Woodward, W.}, \bibinfo{author}{Fuss, C.}, \bibinfo{author}{Ferencik, M.}, \bibinfo{author}{Tereshchenko, L.G.}, \bibinfo{year}{2018}.
\newblock \bibinfo{title}{Torso geometry reconstruction and body surface electrode localization using three-dimensional photography}.
\newblock \bibinfo{journal}{Journal of Electrocardiology} \bibinfo{volume}{51}, \bibinfo{pages}{60--67}.
\bibitem[{Petersen et~al.(2016)Petersen, Matthews, Francis, Robson, Zemrak, Boubertakh, Young, Hudson, Weale, Garratt, Collins, Piechnik and Neubauer}]{journal/JCMR/petersen2016}
\bibinfo{author}{Petersen, S.E.}, \bibinfo{author}{Matthews, P.M.}, \bibinfo{author}{Francis, J.M.}, \bibinfo{author}{Robson, M.D.}, \bibinfo{author}{Zemrak, F.}, \bibinfo{author}{Boubertakh, R.}, \bibinfo{author}{Young, A.A.}, \bibinfo{author}{Hudson, S.}, \bibinfo{author}{Weale, P.}, \bibinfo{author}{Garratt, S.}, \bibinfo{author}{Collins, R.}, \bibinfo{author}{Piechnik, S.}, \bibinfo{author}{Neubauer, S.}, \bibinfo{year}{2016}.
\newblock \bibinfo{title}{Uk biobank's cardiovascular magnetic resonance protocol}.
\newblock \bibinfo{journal}{Journal of cardiovascular magnetic resonance} \bibinfo{volume}{18}, \bibinfo{pages}{8}.
\bibitem[{Pishchulin et~al.(2017)Pishchulin, Wuhrer, Helten, Theobalt and Schiele}]{journal/PR/pishchulin2017}
\bibinfo{author}{Pishchulin, L.}, \bibinfo{author}{Wuhrer, S.}, \bibinfo{author}{Helten, T.}, \bibinfo{author}{Theobalt, C.}, \bibinfo{author}{Schiele, B.}, \bibinfo{year}{2017}.
\newblock \bibinfo{title}{Building statistical shape spaces for {3D} human modeling}.
\newblock \bibinfo{journal}{Pattern Recognition} \bibinfo{volume}{67}, \bibinfo{pages}{276--286}.
\bibitem[{Prakosa et~al.(2018)Prakosa, Arevalo, Deng, Boyle, Nikolov, Ashikaga, Blauer, Ghafoori, Park, Blake~III, Han, MacLeod, Halperin, Callans, Ranjan, Chrispin, Nazarian and Trayanova}]{journal/Nature_BME/prakosa2018}
\bibinfo{author}{Prakosa, A.}, \bibinfo{author}{Arevalo, H.J.}, \bibinfo{author}{Deng, D.}, \bibinfo{author}{Boyle, P.M.}, \bibinfo{author}{Nikolov, P.P.}, \bibinfo{author}{Ashikaga, H.}, \bibinfo{author}{Blauer, J.J.}, \bibinfo{author}{Ghafoori, E.}, \bibinfo{author}{Park, C.J.}, \bibinfo{author}{Blake~III, R.C.}, \bibinfo{author}{Han, F.T.}, \bibinfo{author}{MacLeod, R.S.}, \bibinfo{author}{Halperin, H.R.}, \bibinfo{author}{Callans, D.J.}, \bibinfo{author}{Ranjan, R.}, \bibinfo{author}{Chrispin, J.}, \bibinfo{author}{Nazarian, S.}, \bibinfo{author}{Trayanova, N.A.}, \bibinfo{year}{2018}.
\newblock \bibinfo{title}{Personalized virtual-heart technology for guiding the ablation of infarct-related ventricular tachycardia}.
\newblock \bibinfo{journal}{Nature Biomedical Engineering} \bibinfo{volume}{2}, \bibinfo{pages}{732--740}.
\bibitem[{Qian et~al.(2023)Qian, Ugurlu, Fairweather, Strocchi, Toso, Deng, Plank, Vigmond, Razavi, Young, Lamata, Bishop and Niederer}]{journal/medRxiv/qian2023}
\bibinfo{author}{Qian, S.}, \bibinfo{author}{Ugurlu, D.}, \bibinfo{author}{Fairweather, E.}, \bibinfo{author}{Strocchi, M.}, \bibinfo{author}{Toso, L.D.}, \bibinfo{author}{Deng, Y.}, \bibinfo{author}{Plank, G.}, \bibinfo{author}{Vigmond, E.}, \bibinfo{author}{Razavi, R.}, \bibinfo{author}{Young, A.}, \bibinfo{author}{Lamata, P.}, \bibinfo{author}{Bishop, P.}, \bibinfo{author}{Niederer, S.}, \bibinfo{year}{2023}.
\newblock \bibinfo{title}{Developing cardiac digital twins at scale: Insights from personalised myocardial conduction velocity}.
\newblock \bibinfo{journal}{medRxiv} , \bibinfo{pages}{2023--12}.
\bibitem[{Qiu et~al.(2023)Qiu, Li, Wang, Zhang, Chen, Yang and Zhuang}]{journal/MedIA/qiu2023}
\bibinfo{author}{Qiu, J.}, \bibinfo{author}{Li, L.}, \bibinfo{author}{Wang, S.}, \bibinfo{author}{Zhang, K.}, \bibinfo{author}{Chen, Y.}, \bibinfo{author}{Yang, S.}, \bibinfo{author}{Zhuang, X.}, \bibinfo{year}{2023}.
\newblock \bibinfo{title}{Myo{PS-Net}: Myocardial pathology segmentation with flexible combination of multi-sequence {CMR} images}.
\newblock \bibinfo{journal}{Medical Image Analysis} \bibinfo{volume}{84}, \bibinfo{pages}{102694}.
\bibitem[{Rajaganeshan et~al.(2008)Rajaganeshan, Ludlam, Francis, Parasramka and Sutton}]{journal/JCP/rajaganeshan2008}
\bibinfo{author}{Rajaganeshan, R.}, \bibinfo{author}{Ludlam, C.}, \bibinfo{author}{Francis, D.}, \bibinfo{author}{Parasramka, S.}, \bibinfo{author}{Sutton, R.}, \bibinfo{year}{2008}.
\newblock \bibinfo{title}{Accuracy in {ECG} lead placement among technicians, nurses, general physicians and cardiologists}.
\newblock \bibinfo{journal}{International Journal of Clinical Practice} \bibinfo{volume}{62}, \bibinfo{pages}{65--70}.
\bibitem[{Ramanathan et~al.(2004)Ramanathan, Ghanem, Jia, Ryu and Rudy}]{journal/Nature_Medicine/ramanathan2004}
\bibinfo{author}{Ramanathan, C.}, \bibinfo{author}{Ghanem, R.N.}, \bibinfo{author}{Jia, P.}, \bibinfo{author}{Ryu, K.}, \bibinfo{author}{Rudy, Y.}, \bibinfo{year}{2004}.
\newblock \bibinfo{title}{Noninvasive electrocardiographic imaging for cardiac electrophysiology and arrhythmia}.
\newblock \bibinfo{journal}{Nature Medicine} \bibinfo{volume}{10}, \bibinfo{pages}{422--428}.
\bibitem[{Rjoob et~al.(2020)Rjoob, Bond, Finlay, McGilligan, Leslie, Rababah, Guldenring, Iftikhar, Knoery, McShane and Peace}]{journal/JE/rjoob2020}
\bibinfo{author}{Rjoob, K.}, \bibinfo{author}{Bond, R.}, \bibinfo{author}{Finlay, D.}, \bibinfo{author}{McGilligan, V.}, \bibinfo{author}{Leslie, S.J.}, \bibinfo{author}{Rababah, A.}, \bibinfo{author}{Guldenring, D.}, \bibinfo{author}{Iftikhar, A.}, \bibinfo{author}{Knoery, C.}, \bibinfo{author}{McShane, A.}, \bibinfo{author}{Peace, A.}, \bibinfo{year}{2020}.
\newblock \bibinfo{title}{Machine learning techniques for detecting electrode misplacement and interchanges when recording ecgs: a systematic review and meta-analysis}.
\newblock \bibinfo{journal}{Journal of Electrocardiology} \bibinfo{volume}{62}, \bibinfo{pages}{116--123}.
\bibitem[{Roudijk et~al.(2021)Roudijk, Boonstra, Ruisch, Kastelein, van Dam, Schellenkens, Loh and van Dam}]{journal/JE/roudijk2021}
\bibinfo{author}{Roudijk, R.W.}, \bibinfo{author}{Boonstra, M.J.}, \bibinfo{author}{Ruisch, J.}, \bibinfo{author}{Kastelein, M.}, \bibinfo{author}{van Dam, E.}, \bibinfo{author}{Schellenkens, M.}, \bibinfo{author}{Loh, P.}, \bibinfo{author}{van Dam, P.M.}, \bibinfo{year}{2021}.
\newblock \bibinfo{title}{Feasibility study of a 3d camera to reduce electrode repositioning errors during longitudinal ecg acquisition}.
\newblock \bibinfo{journal}{Journal of Electrocardiology} \bibinfo{volume}{66}, \bibinfo{pages}{69--76}.
\bibitem[{Roy et~al.(2020)Roy, Shah, Villa-Lopez, Murillo, Arenas, Oshima, Chang, Lauzon, Guo and Pillutla}]{journal/JE/roy2020}
\bibinfo{author}{Roy, S.K.}, \bibinfo{author}{Shah, S.U.}, \bibinfo{author}{Villa-Lopez, E.}, \bibinfo{author}{Murillo, M.}, \bibinfo{author}{Arenas, N.}, \bibinfo{author}{Oshima, K.}, \bibinfo{author}{Chang, R.K.}, \bibinfo{author}{Lauzon, M.}, \bibinfo{author}{Guo, X.}, \bibinfo{author}{Pillutla, P.}, \bibinfo{year}{2020}.
\newblock \bibinfo{title}{Comparison of electrocardiogram quality and clinical interpretations using prepositioned {ECG} electrodes and conventional individual electrodes}.
\newblock \bibinfo{journal}{Journal of Electrocardiology} \bibinfo{volume}{59}, \bibinfo{pages}{126--133}.
\bibitem[{Salvador et~al.(2024)Salvador, Kong, Peirlinck, Parker, Chubb, Dubin and Marsden}]{journal/JRSI/salvador2024}
\bibinfo{author}{Salvador, M.}, \bibinfo{author}{Kong, F.}, \bibinfo{author}{Peirlinck, M.}, \bibinfo{author}{Parker, D.W.}, \bibinfo{author}{Chubb, H.}, \bibinfo{author}{Dubin, A.M.}, \bibinfo{author}{Marsden, A.L.}, \bibinfo{year}{2024}.
\newblock \bibinfo{title}{Digital twinning of cardiac electrophysiology for congenital heart disease}.
\newblock \bibinfo{journal}{Journal of the Royal Society Interface} \bibinfo{volume}{21}, \bibinfo{pages}{20230729}.
\bibitem[{Sander et~al.(2023)Sander, de~Vos, Bruns, Planken, Viergever, Leiner and I{\v{s}}gum}]{journal/CBM/sander2023}
\bibinfo{author}{Sander, J.}, \bibinfo{author}{de~Vos, B.D.}, \bibinfo{author}{Bruns, S.}, \bibinfo{author}{Planken, N.}, \bibinfo{author}{Viergever, M.A.}, \bibinfo{author}{Leiner, T.}, \bibinfo{author}{I{\v{s}}gum, I.}, \bibinfo{year}{2023}.
\newblock \bibinfo{title}{Reconstruction and completion of high-resolution {3D} cardiac shapes using anisotropic {CMRI} segmentations and continuous implicit neural representations}.
\newblock \bibinfo{journal}{Computers in Biology and Medicine} \bibinfo{volume}{164}, \bibinfo{pages}{107266}.
\bibitem[{Smith et~al.(2022)Smith, Banerjee, Choudhury and Grau}]{conf/EMBC/smith2022}
\bibinfo{author}{Smith, H.J.}, \bibinfo{author}{Banerjee, A.}, \bibinfo{author}{Choudhury, R.P.}, \bibinfo{author}{Grau, V.}, \bibinfo{year}{2022}.
\newblock \bibinfo{title}{Automated torso contour extraction from clinical cardiac {MR} slices for {3D} torso reconstruction}, in: \bibinfo{booktitle}{IEEE Engineering in Medicine \& Biology Society}, \bibinfo{organization}{IEEE}. pp. \bibinfo{pages}{3809--3813}.
\bibitem[{Smith et~al.(2023)Smith, Rodriguez, Sang, Beetz, Choudhury, Grau and Banerjee}]{journal/arXiv/smith2023}
\bibinfo{author}{Smith, H.J.}, \bibinfo{author}{Rodriguez, B.}, \bibinfo{author}{Sang, Y.}, \bibinfo{author}{Beetz, M.}, \bibinfo{author}{Choudhury, R.}, \bibinfo{author}{Grau, V.}, \bibinfo{author}{Banerjee, A.}, \bibinfo{year}{2023}.
\newblock \bibinfo{title}{Anatomical basis of sex differences in human post-myocardial infarction {ECG} phenotypes identified by novel automated torso-cardiac {3D} reconstruction}.
\newblock \bibinfo{journal}{arXiv preprint arXiv:2312.13976} .
\bibitem[{Tang et~al.(2022)Tang, Gong, Yi, Xie and Ma}]{conf/CVPR/tang2022}
\bibinfo{author}{Tang, J.}, \bibinfo{author}{Gong, Z.}, \bibinfo{author}{Yi, R.}, \bibinfo{author}{Xie, Y.}, \bibinfo{author}{Ma, L.}, \bibinfo{year}{2022}.
\newblock \bibinfo{title}{Lake-net: Topology-aware point cloud completion by localizing aligned keypoints}, in: \bibinfo{booktitle}{Proceedings of the IEEE/CVF Conference on Computer Vision and Pattern Recognition}, pp. \bibinfo{pages}{1726--1735}.
\bibitem[{Tuzcu and Nas(2005)}]{conf/CSMC/tuzcu2005}
\bibinfo{author}{Tuzcu, V.}, \bibinfo{author}{Nas, S.}, \bibinfo{year}{2005}.
\newblock \bibinfo{title}{Dynamic time warping as a novel tool in pattern recognition of {ECG} changes in heart rhythm disturbances}, in: \bibinfo{booktitle}{IEEE International Conference on Systems, Man and Cybernetics}, \bibinfo{organization}{IEEE}. pp. \bibinfo{pages}{182--186}.
\bibitem[{Wan et~al.(2023)Wan, Li, Jia, Gao, Qian, Wu, Lin, Mu, Gao, Wang, Wu and Zhuang}]{journal/MedIA/wan2023}
\bibinfo{author}{Wan, K.}, \bibinfo{author}{Li, L.}, \bibinfo{author}{Jia, D.}, \bibinfo{author}{Gao, S.}, \bibinfo{author}{Qian, W.}, \bibinfo{author}{Wu, Y.}, \bibinfo{author}{Lin, H.}, \bibinfo{author}{Mu, X.}, \bibinfo{author}{Gao, X.}, \bibinfo{author}{Wang, S.}, \bibinfo{author}{Wu, F.}, \bibinfo{author}{Zhuang, X.}, \bibinfo{year}{2023}.
\newblock \bibinfo{title}{Multi-target landmark detection with incomplete images via reinforcement learning and shape prior embedding}.
\newblock \bibinfo{journal}{Medical Image Analysis} , \bibinfo{pages}{102875}.
\bibitem[{Wang et~al.(2020)Wang, Zhou, Zhao, Zhao and Ma}]{conf/JP/wang2020}
\bibinfo{author}{Wang, P.}, \bibinfo{author}{Zhou, Q.}, \bibinfo{author}{Zhao, Y.}, \bibinfo{author}{Zhao, S.}, \bibinfo{author}{Ma, J.}, \bibinfo{year}{2020}.
\newblock \bibinfo{title}{A robust facial landmark detection in uncontrolled natural condition}.
\newblock \bibinfo{journal}{Journal of Physics: Conference Series} \bibinfo{volume}{1693}, \bibinfo{pages}{012202}.
\bibitem[{Yuan et~al.(2018)Yuan, Khot, Held, Mertz and Hebert}]{conf/3DV/yuan2018}
\bibinfo{author}{Yuan, W.}, \bibinfo{author}{Khot, T.}, \bibinfo{author}{Held, D.}, \bibinfo{author}{Mertz, C.}, \bibinfo{author}{Hebert, M.}, \bibinfo{year}{2018}.
\newblock \bibinfo{title}{{PCN}: Point completion network}, in: \bibinfo{booktitle}{2018 International Conference on 3D Vision}, \bibinfo{organization}{IEEE}. pp. \bibinfo{pages}{728--737}.
\bibitem[{Zacur et~al.(2017)Zacur, Minchole, Villard, Carapella, Ariga, Rodriguez and Grau}]{conf/MICCAI/zacur2017}
\bibinfo{author}{Zacur, E.}, \bibinfo{author}{Minchole, A.}, \bibinfo{author}{Villard, B.}, \bibinfo{author}{Carapella, V.}, \bibinfo{author}{Ariga, R.}, \bibinfo{author}{Rodriguez, B.}, \bibinfo{author}{Grau, V.}, \bibinfo{year}{2017}.
\newblock \bibinfo{title}{{MRI}-based heart and torso personalization for computer modeling and simulation of cardiac electrophysiology}, in: \bibinfo{booktitle}{Imaging for Patient-Customized Simulations and Systems for Point-of-Care Ultrasound}, \bibinfo{organization}{Springer}. pp. \bibinfo{pages}{61--70}.
\bibitem[{Zappon et~al.(2024)Zappon, Gsell, Gillette and Plank}]{journal/JE/zappon2024}
\bibinfo{author}{Zappon, E.}, \bibinfo{author}{Gsell, M.A.}, \bibinfo{author}{Gillette, K.}, \bibinfo{author}{Plank, G.}, \bibinfo{year}{2024}.
\newblock \bibinfo{title}{Quantifying variabilities in cardiac digital twin models of the electrocardiogram}.
\newblock \bibinfo{journal}{arXiv preprint arXiv:2407.17146} .
\bibitem[{Zemzemi et~al.(2015)Zemzemi, Dobrzynski, Bear, Potse, Dallet, Coudi{\`e}re, Dubois and Duchateau}]{conf/CinC/zemzemi2015}
\bibinfo{author}{Zemzemi, N.}, \bibinfo{author}{Dobrzynski, C.}, \bibinfo{author}{Bear, L.}, \bibinfo{author}{Potse, M.}, \bibinfo{author}{Dallet, C.}, \bibinfo{author}{Coudi{\`e}re, Y.}, \bibinfo{author}{Dubois, R.}, \bibinfo{author}{Duchateau, J.}, \bibinfo{year}{2015}.
\newblock \bibinfo{title}{Effect of the torso conductivity heterogeneities on the {ECGI} inverse problem solution}, in: \bibinfo{booktitle}{Computing in Cardiology}, \bibinfo{organization}{IEEE}. pp. \bibinfo{pages}{233--236}.
\bibitem[{Zhang et~al.(2021)Zhang, Burrage, Lukaschuk, Shanmuganathan, Popescu, Nikolaidou, Mills, Werys, Hann, Barutcu, Polat, Investigators, Salerno, Jerosch-Herold, Kwong, Watkins, Kramer, Neubauer, Ferreira and Piechnik}]{journal/Circulation/zhang2021}
\bibinfo{author}{Zhang, Q.}, \bibinfo{author}{Burrage, M.K.}, \bibinfo{author}{Lukaschuk, E.}, \bibinfo{author}{Shanmuganathan, M.}, \bibinfo{author}{Popescu, I.A.}, \bibinfo{author}{Nikolaidou, C.}, \bibinfo{author}{Mills, R.}, \bibinfo{author}{Werys, K.}, \bibinfo{author}{Hann, E.}, \bibinfo{author}{Barutcu, A.}, \bibinfo{author}{Polat, S.D.}, \bibinfo{author}{Investigators, H.C.R.H.}, \bibinfo{author}{Salerno, M.}, \bibinfo{author}{Jerosch-Herold, M.}, \bibinfo{author}{Kwong, R.Y.}, \bibinfo{author}{Watkins, H.C.}, \bibinfo{author}{Kramer, C.M.}, \bibinfo{author}{Neubauer, S.}, \bibinfo{author}{Ferreira, V.M.}, \bibinfo{author}{Piechnik, S.K.}, \bibinfo{year}{2021}.
\newblock \bibinfo{title}{Toward replacing late gadolinium enhancement with artificial intelligence virtual native enhancement for gadolinium-free cardiovascular magnetic resonance tissue characterization in hypertrophic cardiomyopathy}.
\newblock \bibinfo{journal}{Circulation} \bibinfo{volume}{144}, \bibinfo{pages}{589--599}.

\end{thebibliography}

\end{document}